\documentclass[prc,aps,tighten,preprint,showpacs]{revtex4}

\usepackage{graphicx}
\usepackage{dcolumn}
\usepackage{bm}
\usepackage{slashed}
\usepackage{float}
\usepackage{amsmath,amssymb}
\usepackage{hyperref}
\usepackage{color}
\usepackage{grffile}
\usepackage{fancybox}
\newcommand{\sh}[1]{#1\hskip -6pt  / }

\begin{document}

\title{Multi-Nucleon Short-Range Correlation Model for Nuclear Spectral Functions: I. Theoretical Framework}

\author{Oswaldo Artiles and Misak M. Sargsian}
\affiliation{Department of Physics, Florida International University, Miami,
Florida 33199, USA}
  
\date{\today}

\begin{abstract}
We develop a theoretical approach for  nuclear  spectral functions at high missing momenta and removal energies  based on the 
multi-nucleon short-range correlation~(SRC) model. The approach is based on the effective Feynman diagrammatic 
method which allows to account for the relativistic effects  important in the SRC domain. 
In addition to two-nucleon SRC with center of mass motion 
we derive also the contribution of  three-nucleon SRCs to the nuclear spectral functions.
The latter is modeled based on the assumption that 3N SRCs are a product of  two sequential short range NN interactions. 
This approach allowed us to express the 3N SRC part of the nuclear spectral function  as a convolution of two NN SRCs.
Thus the knowledge of 2N SRCs allows us to model both two- and three-nucleon SRC contributions to the spectral function.
The derivations of the spectral functions are based on the two  theoretical frameworks in evaluating covariant Feynman diagrams: 
In the first,  referred as virtual nucleon approximation,   we  reduce  Feynman diagrams to the time ordered  noncovariant 
diagrams  by evaluating nucleon spectators in the SRC at their positive energy poles, neglecting  explicitly the contribution 
from vacuum diagrams.
In the second approach, referred as light-front  approximation, we formulate the boost invariant nuclear spectral  
function in the  light-front reference frame in which case the vacuum diagrams are generally suppressed and the 
bound nucleon is described by its light-cone variables such as  momentum  fraction, transverse momentum and invariant mass. 
\end{abstract}

\maketitle

\section{Introduction} 
\label{sec:intro}

The knowledge of the nuclear spectral functions at high momenta  of bound nucleon becomes increasingly 
important for further studies of  nuclear QCD such as medium modification effects (EMC effects) or evolution equation of
partons in nuclear medium measured at very large $Q^2$.     

The importance of the high momentum properties  of bound nucleon for  nuclear EMC effects  follows from the  
recent observations of apparent correlation between medium modification of partonic distributions  and  the strength  
of the two-nucleon short range correlations~(SRCs) in nuclei\cite{LW1,LW2}.   
Concerning  to the QCD evolution of nuclear partonic distributions (PDFs),
one expects that at very large $Q^2$ the knowledge  of high momentum component of 
the nuclear spectral function becomes important due to contribution of   quarks  with 
momentum fractions  larger than the ones  provided by an isolated nucleon (i.e. partons with $x>1$)\cite{FSS15,FS15}.
The same is true for the reliable interpretation of neutrino-nuclei scatterings in which case both medium modification of 
PDFs as well as realistic treatment of SRCs are essential\cite{Nutev,mNutev}.
All these requires a reasonably well understanding of the nuclear spectral functions   at high  momenta and removal energies
of  bound nucleon. With the advent of the Large Hadron Collider and expected construction of electron-ion colliders as well 
as several ongoing neutrino-nuclei  experiments the knowledge of such  spectral functions will be an important part of the 
theoretical interpretation of the data involving nuclear targets.

Despite  impressive recent progress in ''ab-initio" calculations of nuclear wave functions  their
relevance to the  development of the   spectral functions at large momenta and removal energies is rather limited. 
Not only the absence of  relativistic effects but also the impossibility of identifying  the  relevant $NN$ interaction potentials 
makes such a program unrealistic.  One way of progress  is to  develop   theoretical models based on 
the short range NN correlation approach  in the description of the high momentum part of the nuclear wave function 
(see. e.g. \cite{FS81,Ciofi1991,CiofiSimula,Jan2012,Alvioli2012,Alvioli2013,Rios2013,Jan2014,Ciofi2015,Neff2015}).  
In such approach one  will be able to take into account the empirical knowledge  of SRCs acquired from  different high energy 
scattering experiments thus  reducing in some degree  the theoretical uncertainty related to the description of 
the high momentum nucleon  in the nucleus. 

Our current work is such an attempt, which is  based on the  several   phenomenological observations  
obtained in  recent years in studies of  the  properties on two-nucleon(2N)  SRCs\cite{Kim1,Kim2,isosrc,Eip3,srcrev,Fomin2011,srcprog,Cosyn2015,Eip4,contact}. 
We first develop the model describing nuclear spectral function at large momenta and missing energies 
dominated by 2N SRCs with their center of mass motion generated by the mean field of the A-2 residual nuclear system.
We then develop a theoretical framework for calculating the contribution of  three-nucleon SRCs
to the nuclear spectral function based on the model in which such correlations are generated by sequential short 
range NN interactions. As a result in our approach the  phenomenological knowledge of the properties of NN SRCs is sufficient to 
calculate both 2N and 3N  SRC contribution to the nuclear spectral function.

In Sec.\ref{Sec.2} we give a brief summary of recent advances in studies of the structure of NN SRCs which provides us with 
the phenomenology  for developing the 2N and 3N SRC models of nuclear spectral functions.

Since the domain of multi-nucleon SRCs is characterized by relativistic momentum of the probed  nucleon, 
special care should be given to the treatment of  relativistic effects. 
To identify the relativistic effects,  in Sec.\ref{Sec.3},  we first formulate the nuclear spectral function as 
a quantity which is  extracted in the semi-exclusive  high energy process 
whose scattering amplitude can be described through the covariant effective Feynman diagrams.  
The covariance here is important since it allows consistently trace the relativistic effects related to the propagation of 
the bound nucleon. We then identify the part of the covariant diagram  which reproduces  the nuclear spectral function.  
Doing so, we adopt two approaches for modeling the nuclear spectral function: 
Virtual Nucleon  and Light-Front, general features of which are described in Sec.\ref{Sec.3}. 
Sec.\ref{Sec.4} outlines the calculation of nuclear spectral functions  based on the effective Feynman diagrammatic  method, identifying diagrams 
corresponding to  the mean-field, 2N-SRC with center of mass motion and  3N- SRC contributions.

In  Secs.\ref{Sec.5} and \ref{Sec.6} we  present the detailed derivation of the spectral functions within virtual nucleon and light-cone approximations.
Sec.\ref{Sec.7} summarizes the results. This work represents the theoretical foundation and derivation of spectral functions,  the 
follow-up paper\cite{multisrcII} will present numerical estimates and parameterizations that can be used in practical calculations of 
different nuclear processes.

 \section{Phenomenology of Two Nucleon Short Range Correlations in Nuclei}
 \label{Sec.2}
 
Recent experimental studies of  high energy $eA$ and $pA$ processes\cite{Kim1,Kim2,Fomin2011,isosrc,Eip3,Eip4} 
resulted in a  significant progress in understanding the dynamics  of 2N SRCs in nuclei. 
The series of  electron-nucleus inclusive scattering experiments\cite{Kim1,Kim2,Fomin2011} confirmed the 
prediction\cite{FS88,FSDS93} of the  scaling for the ratios of inclusive cross section of a nucleus 
to the deuteron cross section  in the kinematic region dominated by 
the scattering from the bound nucleons with momenta $p > k_{F}\sim 250$~MeV/c.  Within the 2N SRC model, 
these ratios allowed to extract the parameter $a_2(A,Z)$ which characterizes the probability of finding 2N SRC in 
the nucleus relative to the deuteron.

High energy  semi-inclusive experiments\cite{isosrc,Eip3}  allowed for the first time to probe the isospin composition of 
the 2N SRCs, observing strong  (by factor of 20) dominance  of the  $pn$ SRCs in nuclei,   
as compared to  the $pp$ and $nn$  correlations, for internal momentum range  of $\sim 250-650$~MeV/c. 
This observation is understood\cite{isosrc,eheppn2,Sch} based on the dominance of the tensor forces in the NN interaction at this 
momentum range corresponding to the average nucleon separations of $\sim 1.1$~fm.   
The tensor interaction projects the NN SRC part of the 
wave function to  the  isosinglet - relative angular momentum, $L=2$,  state,  almost identical to the high momentum part of the  $D$-wave component of 
the deuteron wave function.  As a result $pp$ and $nn$ components of the NN SRC are strongly suppressed since they 
are dominated by the  central NN potential with relative angular momentum $L=0$.

Based on the observation of the strong dominance of $pn$ SRCs in Ref.\cite{newprops,proa2} it  was  predicted 
that single proton or neutron  momentum distributions in the 2N SRC domain are inverse proportional to their 
relative fractions in  nuclei.  This prediction is in agreement with the results of  variational Monte-Carlo calculation
of momentum distributions of  light nuclei\cite{Wir2014} as well 
as for medium to heavy nuclei based on  the  SRC model calculations of Ref.\cite{Jan2014}.  
The recent finding  of the $pn$ dominance in heavy nuclei (up to $^{208}Pb$)\cite{Eip4} 
validates the universality of the above prediction for  the whole spectrum of atomic nuclei.
The inverse proportionality of the high momentum component  to the relative fraction of 
the proton or neutron is  important for asymmetric nuclei and they need to be included in the 
modeling of nuclear spectral functions in the 2N SRC region.

Finally, the $pn$ dominance in the SRC region  and its relation to the high momentum part of the deuteron wave function makes 
the studies of the deuteron structure at large internal momenta a very important part for the SRC studies in nuclei.
In this respect the recent experiments\cite{Boeglin11,BS15} and planned new measurements\cite{Boeglinproposal}  of high energy 
exclusive electro-disintegration of the deuteron opens up a new  possibilities in the extraction of the deuteron momentum distribution 
at very large momenta. The measured  distributions ten  can be  utilized in the calculation of the nuclear spectral functions in the 
multi-nucleon SRC region.

 \section{Formulation of Nuclear Spectral Function}
 \label{Sec.3}
 
 Our approach in the definition of nuclear spectral functions is based on  identifying  a nuclear ``observable" which can be 
 extracted from the cross section of  the large momentum ($\gg m_N$)  transfer  semi-inclusive 
 $h+A \rightarrow h^\prime + N + (A-1)^*$ reaction in which the $N$ can be unambiguously identified as a struck nucleon carrying almost 
 all the energy and momentum transferred to the nucleus by the probe $h$. 
The reaction is specifically chosen to be semi-inclusive  that allows, in the approximation in 
which no final state interactions are considered,  to relate  the missing momentum  and energy  of the reaction  to the properties 
of bound nucleon in the nucleus.  With above conditions satisfied the extracted ``observable", referred as a nuclear 
spectral function, represents a joint probability of finding bound nucleon in the nucleus with given missing momentum and energy.
 
In Ref.\cite{gea,ms01} we developed  an effective Feynman 
diagrammatic approach for calculation of the $h+A \rightarrow h^\prime + N + (A-1)^*$ reactions. In this approach the 
covariant Feynman scattering amplitude is expressed through the effective nuclear vertices,  vertices related to the scattering 
of the probe $h$  with the bound nucleon, as well as vertices related to the final state NN interactions. 
The nuclear vertices with the propagator of bound nucleon can not be associated  apriory 
with the single nucleon  wave function of the nucleus, since they  contain negative energy components 
which are related to the  vacuum fluctuations rather than the  probability amplitude of finding nucleon with given momentum 
in the nucleus.
\begin{figure}[h]
\centering\includegraphics[scale=0.7]{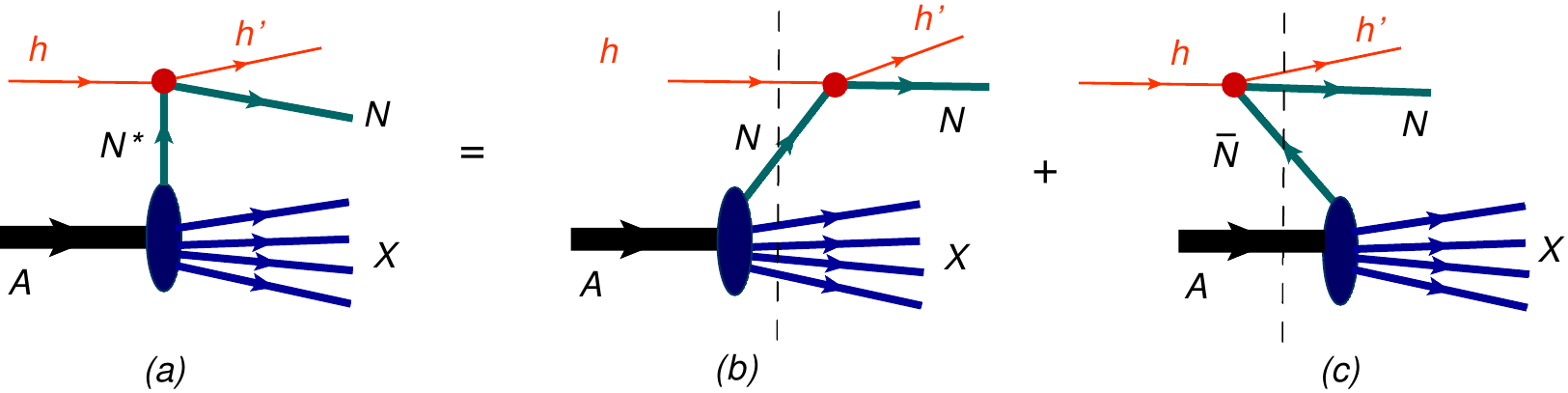}
\caption{Representation of the covariant Feynman amplitude through the sum of the 
time ordered amplitudes.}
\label{Fig:R_diagram}
\end{figure}
This problem is illustrated in the diagrammatic representation of the reaction shown in Fig.\ref{Fig:R_diagram},  in which
the covariant diagram~(a) is  a sum of two non-convariant time ordered scattering diagrams (b) and (c).
Here, for the  calculation of the Lorentz invariant 
amplitude  of Fig.\ref{Fig:R_diagram}(a) one can use the  Feynman diagrammatic rules given in Ref.\cite{ms01}. 
However the nuclear  spectral function can only be formulated for the diagram of Fig.\ref{Fig:R_diagram}(b), where 
the time ordering is such that it first exposes the nucleus as being composed of a  bound nucleon and residual nucleus, 
followed by an  interaction of the incoming probe $h$ off the bound nucleon.  The 
other time ordering  (Fig.\ref{Fig:R_diagram}({c})) presents a  very different scenario of the scattering 
in which the probe produces a $\bar N N$ pair with subsequent absorption of the $\bar N$ in the nucleus. The 
later is usually referred as a $Z$-graph and is not related to the nuclear spectral function. 
It is worth noting that the $Z$-graph contribution is a purely relativistic effect and does not appear in the 
non-relativistic formulation of the nuclear spectral function. Its contribution however increases 
with an increase of the momentum of the bound nucleon (see e.g. Ref.\cite{FS81}). 

The above discussion  indicates that while defining nuclear spectral function is straightforward in the non-relativistic domain 
(no $Z$-graph contribution), its definition becomes increasingly ambiguous  with an  increase of 
bound nucleon momentum.   This ambiguity is reflected in 
the lack of uniqueness in defining the nuclear spectral function in the domain where one expects 
to  probe SRCs.
  
 In the present work we consider two approaches in defining  the nuclear spectral function from   the 
 covariant scattering amplitude.
In the first approach we neglect by the $Z$-graph contribution considering only the positive energy pole for 
the nucleon propagators in the nucleus.  The energy and momentum conservation  in this case  requires the interacting 
nucleon to be virtual which renders certain ambiguity in treating the propagator of the bound nucleon.  The approach we 
follow is to recover the energy and momentum of the interacting nucleon from kinematic parameters of on-shell spectators 
(see Ref.\cite{Gross:1982} for general discussion of the spectator model of relativistic bound states.)
The advantage of this approximation is that the spectral function is expressed through the 
nuclear  wave function defined in the rest frame of the nucleus which in principle can be calculated using the 
conventional NN potentials. We will refer to this approach  as virtual-nucleon~(VN) approximation\cite{noredepn,edepnx}.

In the second approach the nuclear spectral function is defined on the light front which corresponds to a
reference frame in which the nucleus has infinite momentum.   In this  approach,  refereed as Light-Front (LF) approximation, 
the $Z$-graph contribution is kinematically  suppressed\cite{goodcomp} and as a result the invariant sum of 
the two light-cone time ordered amplitudes in Fig.\ref{Fig:R_diagram} is equal to  the  contribution from 
the graph of Fig.\ref{Fig:R_diagram} (b) only (see e.g. Refs\cite{Weinberg,FS81,FS92,GM00}).  
This situation allows us to define boost invariant LF spectral function  in  which the probed nucleon is the constituent of the nucleus 
with given light-cone momentum fraction, transverse momentum and invariant mass. Our approach is field-theoretical 
in which  Feynman diagrams are constructed with effective interaction vertices and the spectral functions are extracted from 
the imaginary part of the covariant forward scattering nuclear amplitude.    Another approach in LF approximation is the 
construction of the  nuclear spectral function based on the relativistic hamiltonian dynamics representing the interaction 
of fixed number on-mass shell constituents\cite{Salme}.  


\section{Diagrammatic Method of Calculation of Nuclear Spectral Function}
\label{Sec.4}

In both VN and LF approximations we can use  the diagrammatic approach of Ref.\cite{ms01}  to 
calculate the spectral functions.  For this we identify the effective interaction vertices  $\hat V$ such 
that imaginary part of the covariant forward scattering  nuclear amplitude will reduce to the nuclear spectral function either in 
VN or LF approximations. 

\begin{figure}[h]
\centering\includegraphics[scale=0.95]{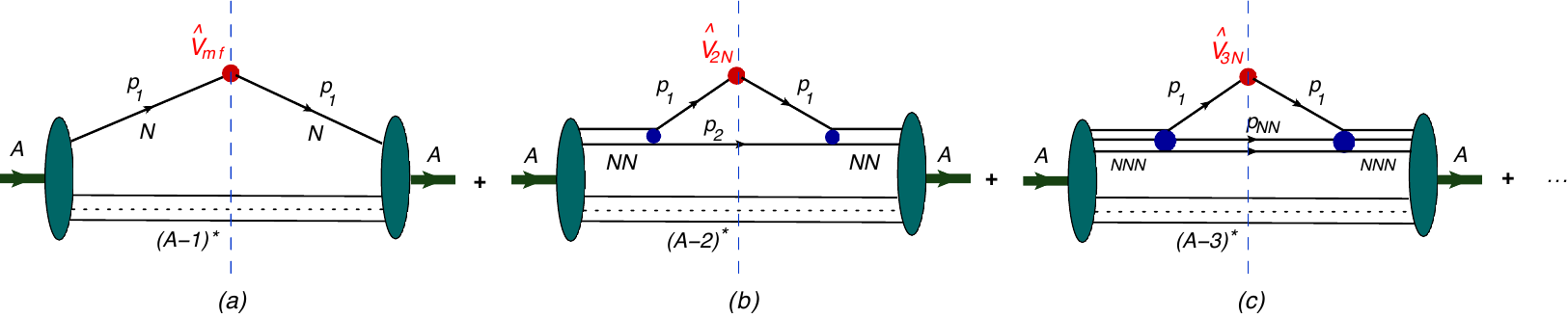}
\caption{Expansion of the nuclear spectral function into the contributions of mean field (a),
2N (b) and 3N ({c}) SRCs.}
\label{Fig:spectral_diagram}
\end{figure}

In applying  the diagrammatic approach  one can express the forward  nuclear scattering amplitude as a sum of  
the   mean-field and multinucleon SRC contributions  as presented in Fig.(\ref{Fig:spectral_diagram}), 
with (a), (b) and (c) corresponding to the contributions from mean-field, 2N and 3N short-range correlations.

Since the mean-field contribution (Fig.(\ref{Fig:spectral_diagram})(a)) is dominated by the momenta of interacting nucleon below 
the characteristic Fermi momentum, $k_{F}$, one can approximate  the corresponding spectral function to the result following 
from non-relativistic calculation.  In this case both VN and LF approximations are expected to give 
very close results.

For 2N SRCs  (Fig.(\ref{Fig:spectral_diagram})(b)), the  momenta of probed nucleon  is $k_{F} <p\le  600-700$~MeV/c  
and the non-relativistic approximation is  increasingly  invalid.
Currently there is a rather robust   phenomenology on 2N SRCs in nuclei\cite{Kim1,Kim2,Fomin2011,isosrc,Eip3,Eip4,srcprog}, which should be taken into account in  the  calculation of  2N SRC  contribution to the nuclear spectral function.

Finally, (Fig.(\ref{Fig:spectral_diagram})(c)), corresponds  to  3N SRCs.   
Currently, there are few rather contradictory  experimental evidences  on  3N SRCs\cite{Kim2, Fomin2011,OrDouglas}
and the  first  high-energy dedicated  studies  are expected in the near future\cite{xg3prop}.
In the present work  we develop a model for  3N SRC which  is based on the assumption that 
3N SRCs are a result of the sequential   short range NN interactions. The final result  represents a convolution of two 2N SRCs.
In this way just the knowledge of 2N SRCs will be sufficient to account for both two- and three-nucleon short range correlation contributions
to the nuclear spectral function.

In the calculation  we apply the effective Feynman rules\cite{ms01} to the covariant  forward scattering 
amplitudes  corresponding to 
mean-field, two- and three-nucleon SRC contributions (Fig.\ref{Fig:spectral_diagram}) separately.  Then within VN or LF approximation  we 
estimate the loop-integrals  through the on-mass shell conditions of intermediate states.   Doing so  we absorb 
nuclear to nucleon transition vertices into the definition of   nuclear wave functions.
In the following calculations we express  the  covariant forward scattering amplitude, $A$,  in the following form:
\begin{equation}
A = A^{MF} + A^{2N} + A^{3N} + \cdots, 
\end{equation}
where   $A^{MF}$ , $A^{2N}$ and $A^{3N}$  correspond to the contributions  from the 
diagrams of  Fig.\ref{Fig:spectral_diagram} (a),(b) and (c) and then consider  each contribution separately.

\subsection{Mean Field Contribution}
\label{sec4a}

In the mean field approximation the probed nucleon interacts with  the nuclear field induced by the  $A-1$ residual system.
In such approximation the  spectral function corresponds to the nuclear configuration in which 
the residual nuclear system is identified as a coherent $A-1$ state with  excitation energy in the order of 
{\em tens} of MeV.

Applying the effective Feynman rules to the diagram of  Fig.\ref{Fig:spectral_diagram}(a) corresponding 
to the mean field contribution of nuclear spectral function one obtains:
\begin{eqnarray}
A^{MF}   =    -Im\int \chi^\dagger_A \Gamma_{A\rightarrow N,A-1}^\dagger 
{{\sh p}_1 + M_N\over p_1^2 - M_N^2 }\hat V^{MF} {\sh p_1 + M_N\over p_1^2 - m^2}
\left[ {G_{A-1}(p_{A-1})\over p_{A-1}^2 - M_{A-1}^2 + i\varepsilon}\right]^{on} \Gamma_{A\rightarrow N,A-1} \chi_A \nonumber \\
\times {d^4p_{A-1}\over i (2\pi)^4},
 \label{eq:Smf}
\end{eqnarray}
where  $M_N$ and $M_{A-1}$ are the masses of nucleon and residual $A-1$ nuclear system,  $\chi_A$ is the nuclear spin wave function, 
$\Gamma_{A\rightarrow  N,A-1}$ represents the covariant vertex of
$A\rightarrow N + A-1$ transition and $G_{A-1}$ describes the propagation of $A-1$ residual nucleus in the 
intermediate state. The  label $[\cdots]^{on}$ indicates that  one estimates the cut diagram in which the residual nuclear system is on mass shell.

\subsection{Two Nucleon  SRC  Contribution} 
\label{sec4b}
In two nucleon SRC model one assumes that the intermediate nuclear state  consists of two correlated  fast ($> k_{F}$)  nucleons and slow  ($< k_{F}$) 
coherent $A-2$ nuclear system.   The corresponding Feynman diagram is presented in Fig.\ref{Fig:spectral_diagram}(b),
for which using the same diagrammatic rules\cite{ms01}  one obtains:
\begin{eqnarray}
A^{2N}  &=  &  Im\int \chi^\dagger_A \Gamma_{A\rightarrow NN,A-2}^\dagger {G(p_{NN})\over p_{NN}^2 - M_{NN}^2 }
 \Gamma^\dagger_{NN\rightarrow NN} {\sh p_1 + M_N\over p_1^2 - M_N^2} \hat V^{2N}
 {\sh p_1 + M_N\over p_1^2 - M_N^2 } 
 \left[ {\sh p_2 + M_N\over p_2^2 - M_N^2 + i\varepsilon}\right]^{on}    \nonumber \\
& & \ \  \times \ \  \Gamma_{NN\rightarrow NN}
 {G(p_{NN})\over p_{NN}^2 - M_{NN}^2 } 
\left[ {G_{A-2}(p_{A-2})\over p_{A-2}^2 - M_{A-2}^2 + i\varepsilon}\right]^{on} \Gamma_{A\rightarrow NN,A-2} \chi_A  
 {d^4p_{2}\over i (2\pi)^4} {d^4p_{A-2}\over i (2\pi)^4}, 
  \label{sp:2N_cov}
 \end{eqnarray}
 where  $M_{NN}$ is the mass of the 2N SRC system, $\Gamma_{A\rightarrow NN,A-2}$ now describes the  transition of the 
 nucleus A to  the  $NN$ SRC and  coherent $A-2$  residual state, while 
 the  $\Gamma_{NN\rightarrow NN}$  vertex describes the short range $NN$ interaction that generates  two-nucleon correlation in the 
 spectral function.

\subsection{Three Nucleon SRC  Contribution in Collinear Approximation} 
\label{sec4c}
The spectral function due to 3N short-range correlations is described in Fig.\ref{Fig:spectral_diagram}({c}) in which the 
intermediate state consists of three fast ($> k_F$) nucleons and slow coherent $A-3$ residual  system.
The dynamics of 3N SRCs allow more complex interactions than that of 2N SRCs.  One of the complexities is 
the irreducible three-nucleon forces that can not be described by NN interaction only.  
Such interactions may contain 
inelastic transitions such as $NN\rightarrow N\Delta$.    As our early studies demonstrate\cite{eheppn2,eheppn1} 
irreducible three-nucleon forces predominantly contribute at very large magnitudes of missing energy characteristic to 
the $\Delta$ excitations $\sim 300$~MeV/c.  Thus for spectral functions at  not very large missing energies one can 
restrict by contributions of  $NN\rightarrow NN$  interactions only.
\begin{figure}[h]
\centering\includegraphics[scale=0.7]{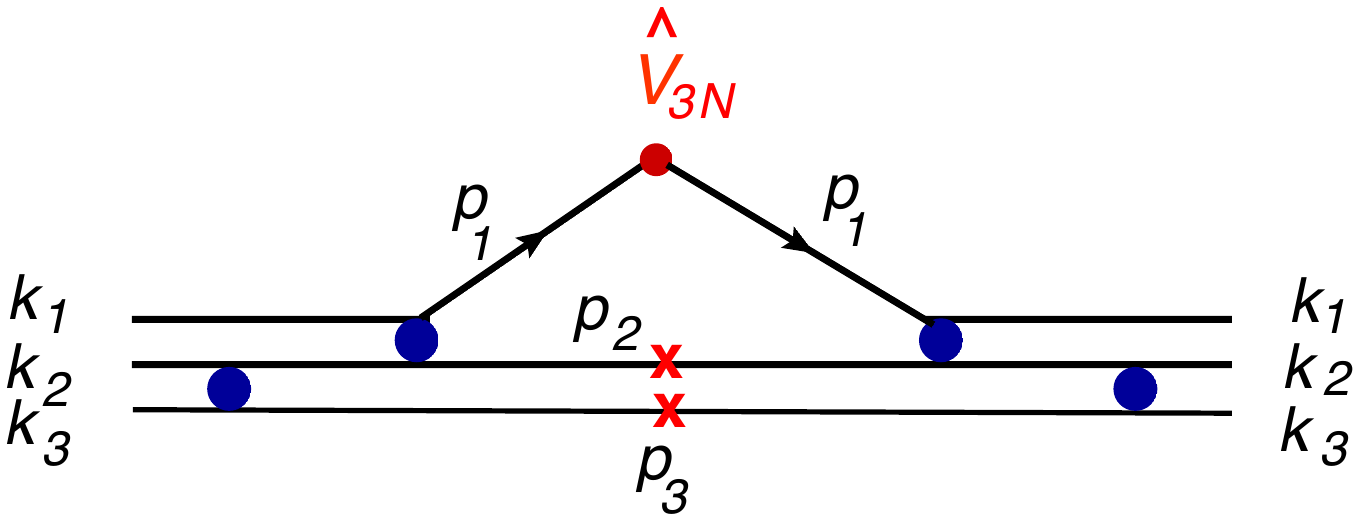}
\caption{Diagram corresponding to 3N SRC  contribution to the spectral function.}
\label{Fig:3nSpectral}
\end{figure}
In the present work  we will follow  the  two  sequential NN short-range interaction  scenario of the  generation of 3N SRCs.  
In this  approximation,   the 3N SRC contribution to   the  spectral function can 
be represented through the diagram of Fig.\ref{Fig:3nSpectral}.   Here  we  factored out  the 
low momentum  residual $A-3$ system from the consideration. This is justified by the fact that  much larger momenta are  involved in the 3N SRCs as compared
to the one in  the 2N SRCs discussed in the previous section. As a result the effects due to center of mass motion of  the $A-3$ system are neglected.
The present approximation assumes that initial three collinear nucleons 
undergo two short range NN interactions generating one nucleon with much larger momenta  than the other 
two. Note that the momenta of all three nucleons in the intermediate state of the scattering in Fig.\ref{Fig:3nSpectral}
should exceed the nuclear Fermi momentum $k_F$ to 
satisfy the 3-particle-3-hole condition in the Fermi distribution of nucleons in the nucleus.

Using the effective Feynman diagrammatic rules for the diagram of Fig.\ref{Fig:3nSpectral} one obtains:
\begin{eqnarray}
 A^{3N} \vspace{-0.9cm}    &  = &  Im\int \bar u(k_1,\lambda_1) \bar u(k_2,\lambda_2) \bar u (k_3,\lambda_3) 
  \Gamma^\dagger_{NN\rightarrow NN} {\sh p_{2^\prime} +M_N  \over p_{2^\prime}^2 - M_N^2}
 \Gamma^\dagger_{NN\rightarrow NN} 
 {\sh p_1 + M_N\over p_1^2 - M_N^2} \hat V^{3N}
 {\sh p_1 + M_N\over p_1^2 - M_N^2}  \nonumber  \\
& &  \times \  \left[  {\sh p_2 + M_N\over p_2^2 - M_N^2 + i\varepsilon} \right]^{on}\Gamma_{NN\rightarrow NN}
{\sh p_{2^\prime} + M_N\over p_{2^\prime}^2 - M_N^2} 
\left[ {\sh p_{3} + M_N\over p_{3}^2 - M_N^2 + i\varepsilon}\right]^{on}  \Gamma_{NN\rightarrow NN}  \nonumber \\
& & \times  \ u(k_1,\lambda_1)  u(k_2,\lambda_2)  u(k_3,\lambda_3) 
  {d^4p_{3}\over i (2\pi)^4} {d^4p_{2}\over i (2\pi)^4}, 
 \label{sp:3N_cov}
 \end{eqnarray}
where ``$2^\prime$" labels the intermediate state of the nucleon ``2" after the first short-range NN interaction, 
$\lambda_i$ is the spin of the $i$'th nucleon 
and the $\Gamma_{NN\rightarrow NN}$ is the  same short range NN interaction vertex  included in  Eq.(\ref{sp:2N_cov}).

\subsection{Models of Calculation}
\label{sec4d}
To calculate the spectral functions from the forward scattering amplitudes in
Eqs.(\ref{eq:Smf}), (\ref{sp:2N_cov}) and (\ref{sp:3N_cov}) 
one needs to define
the effective vertices  $\hat V$ which identify the bound nucleon in the mean field, 2N and 3N SRCs, as well as to define the poles
at which  the cut propagators of the intermediate states are  estimated.
Both depend on the approximation used to reduce the covariant diagrams to the time ordered diagrams which allow 
an introduction of the  nuclear spectral function.
In the following section we will derive these spectral functions from the covariant forward scattering amplitudes $A^{MF}$, $A^{2N}$ and $A^{3N}$ within VN and LF 
approximations.

 \section{Spectral Function in Virtual Nucleon Approximation}
\label{Sec.5}

 Our first approach is VN approximation which describes the nucleus in the Lab frame treating interacting bound nucleon as a virtual particle  
 while  spectators being on their mass-shells.  In VN  approximation the spectral function, $S^{N}_A(p,E_m)$ defines the joint probability of finding  a nucleon in the nucleus  with momentum $p$ and removal energy $E_m$ defined as
 \begin{equation}
 E_m = E_{A-1} + M_N - M_{A} - {p^2\over 2M_{A-1}},
\label{EM}
\end{equation}
where $E_{A-1}$  and $M_{A-1}$ are the energy and the mass of the residual $A-1$ nuclear system.
  The normalization condition for the spectral function can be fixed from the condition of the conservation of baryonic number of the 
  nucleus in hadron-nucleus scattering\cite{FS87} or from the condition for the charge form-factor of nucleus  at vanishing momentum transfer, 
  $F_A(0)= Z$\cite{noredepn} which yields:
\begin{equation}
\sum\limits_N\int  S^{N}_A(p,E_m)  \alpha d^3p  dE_m    = 1.
\label{bnorm}
\end{equation}
where $\alpha$ is  the ratio of the flux factors of the (external probe) -(bound nucleon) and  (external probe)-(nucleus) systems, 
which in high momentum limit of  the probe ( hadron or virtual photon) yields
\begin{equation}
\alpha = {E_N + p_{z} \over M_A/A}.
\label{alpha}
\end{equation}
Here $E_N$ is the energy of the bound nucleon and the $z$ direction is defined opposite to the direction of 
incoming probe.

Following the decomposition of Fig.\ref{Fig:spectral_diagram} we consider the  mean-field, 2N and 3N SRC contributions 
to the nuclear spectral function separately. 
In VN approximation the cut diagrams of Fig.\ref{Fig:spectral_diagram} and Fig.\ref{Fig:3nSpectral} will be evaluated at positive energy poles of the spectator residual system. 
For the mean field contribution it  corresponds to the positive energy pole of the A-1 system, for the 2N  and 3N SRCs these  
are the positive energy poles of correlated one and two nucleons respectively.

\subsection{Mean Field Contribution}
\label{sec5a}
 
In the mean-field approximation the  missing momentum ${\bf p_m} = - {\bf p} \equiv - {\bf p_1}$ and missing energy $E_m$ 
characterizes the total momentum and  excitation energy 
of the residual $A-1$ system.  In the nuclear -shell model, $E_m$ also defines  the energy needed to remove the 
nucleon from the particular nuclear shell. 
For such a situation we can define the effective vertex $\hat V_{MF}$ in Eq.(\ref{eq:Smf}) as 
\begin{equation}
\hat V^{MF}  = i \bar a(p_1,s_1) \delta^3({\bf p_1} + {\bf p_{A-1}})\delta(E_m - E_\alpha) a(p_1,s_1),
\end{equation}
where $E_\alpha$ is the characteristic energy of the given nuclear shell. The creation and annihilation operators 
are defined in such a way that:
\begin{equation}
a(p_1,s_1)(\sh p_1 + M_N) = \bar u(p_1,s_1) \ \ \ \mbox{and}  \ \ \  (\sh p_1 + M_N) \bar a(p_1,s_1,t_1) =  u(p_1,s_1).
\label{aops}
\end{equation}
Next, in Eq.(\ref{eq:Smf}) we take the integral by $d^0 p_{A-1}$    through the positive energy pole of the 
propagator of the $A-1$ state:
\begin{equation}
{dp^0_{A-1}\over p_{A-1}^2 - M_{A-1}(E_{\alpha})^2 + i\varepsilon}   =   - {2\pi i\over 2 E_{A-1}}\left |_{E_{A-1} = \sqrt{M_{A-1}^2  + p_{A-1}^2}}\right.,
\end{equation}
and for the on-shell (A-1) spectator state  we use the sum rule:
\begin{equation}
G_{A-1}(p_{A-1}) = \sum\limits_{s_{A-1}} \chi_{A-1}(p_{A-1},s_{A-1},E_\alpha) \chi^\dagger_{A-1}(p_{A-1},s_{A-1},E_\alpha),
\end{equation}
where $\chi_{A-1}$ is the spin wave function of the residual (A-1) nucleus.
The above relations allow us to introduce the single nucleon wave function for the given nuclear shell $E_\alpha$, 
$\psi_{N/A}$   in the form:
\begin{equation}
\psi_{N/A}(p_1,s_1,p_{A-1},s_{A-1},E_\alpha) = { \bar u(p_1,s_1)\chi^\dagger_{A-1}(p_{A-1},s_{A-1},E_\alpha)\Gamma_{A,N,A-1} \chi_A\over
(M_{N}^2 - p_{1}^2) \sqrt{(2\pi)^3 2E_{A-1}}},
\end{equation}
which, inserting into Eq.(\ref{eq:Smf}) and taking the $d^3p_{A-1}$ integration through the $\delta^3(p_1+p_{A-1})$,
results in the mean-field nuclear spectral function, $S^{N}_{A,MF}$ in the VN approximation::
\begin{equation}
S^{N}_{A,MF} (p_1,E_m) = \sum\limits_{\alpha}\sum\limits_{s_1,s_{A-1}} \mid 
\psi_{N/A}(p_1,s_1,p_{A-1},s_{A-1},E_\alpha) \mid^2 \delta(E_m - E_\alpha)
\end{equation}
For numerical estimates of the above 
spectral function we  note that  in the mean-field approximation, the substantial strength of the VN  wave function 
$\psi_{N/A}$  comes from the momentum range of $p_1\le k_{F}$. Therefore in this case 
the nonrelativistic approximation is valid,  which allows us to approximate this wave function by the nonrelativistic wave function 
obtained  from the conventional mean field calculations of single nucleon wave functions. Additionally, in the non-relativistic limit
$\alpha \approx 1 + {p_{1,z}\over M_A/A}$, and from 
Eq.(\ref{bnorm}) one observes that the  ${p_{1,z}\over M_A/A}$ part does not contribute to the integral resulting 
to the condition for non-relativistic normalization:
\begin{equation}
\int  S^{N}_{A,MF}(p,E_m) dE_m d^3p     = n^N_{MF},
\label{nrnorm}
\end{equation}
where $n^N_{MF}$ is the  norm of the  mean field contribution of nucleon N
to the total normalization of the nuclear spectral function.

\subsection{Two Nucleon Short-Range Correlations}
\label{sec5b}
We consider now Eq.(\ref{sp:2N_cov}) with the effective vertex $\hat V_{2N}$ identifed as
\begin{equation}
\hat V_{2N}  = i  \bar a(p_1,s_1) \delta^3({\bf p_1} + {\bf p_2} + {\bf p_{A-2}})\delta(E_m - E_m^{2N}) a (p_1,s_1),
\end{equation}
where   $\bar a(p_1,s_1,t_1)$ and $a(p_1,s_1,t_1)$ are creation and annihilation operators of nucleon with 
four-momentum $p_1$ and spin $s_1$ satisfying the relations of Eq.(\ref{aops}).
 
The magnitude of $E_m^{2N}$ is follows from the NN correlation model (in which the correlated NN pair has a 
total momentum $-{\bf p_{A-2}}$  in the mean field of $A-2$ residual nuclear system)  according to which:
\begin{equation}
E_m^{2N} = E^{(2)}_{thr} + T_{A-2} + T_{2} - T_{A-1}=   E^{(2)}_{thr} + {p_{A-2}^2\over 2M_{A-2}} +
T_2  - {p_{1}^2\over 2 M_{A-1}},
\end{equation}
where $E^{(2)}_{thr}$ is the threshold energy needed to remove two nucleons from the nucleus, 
$T_2$, $T_{A-2}$ and $T_{A-1}$ are kinetic energies of  the correlated nucleon, ``2",  
residual $(A-2)$  nucleus and $(A-1)$ nuclear system (consisting of the ``2" nucleon and (A-2) nucleus).
Here the kinematic energies $T_{A-2}$ and $T_{A-1}$ are treated nonrelativistically.

According to VN prescription we perform integrations by $dp^0_2$ and $dp^0_{A-2}$  in   Eq.(\ref{sp:2N_cov})  through the
positive energy poles of the  propagators of $"2"$ and $"A-2"$ particles. This yields
\begin{eqnarray}
{dp^0_2\over p_2^2 - M_N^2 + i\varepsilon}  & =  & - {2\pi i\over 2E_2}\left |_{E_2 = \sqrt{M_N^2  + p_2^2}}\right. \nonumber \\
{dp^0_{A-2}\over p_{A-2}^2 - M_{A-2}^2 + i\varepsilon}  & =  & - {2\pi i\over 2E_{A-2}}\left |_{E_{A-2} = \sqrt{M_{A-2}^2  + p_{A-2}^2}}\right. 
\label{vn_poles}
\end{eqnarray}
Since ``2" and ``A-2" are now on mass shell, we can write the sum rule relations for the numerator of their propagators as:
\begin{eqnarray}
\sh p_2 + M_N & = &  \sum\limits_{s2}u(p_2,s_2)\bar u(p_2,s_2) \\ \nonumber
 G(p_{A-2})  & = &  \sum\limits_{s_{A-2}}\chi_{A-2}(p_{A-2},s_{A-2})\chi^\dagger_{A-2}(p_{A-2},s_{A-2}),
\label{closurule}
\end{eqnarray}
where $\chi_{A-2}$ is the spin wave function of the $A-2$ nucleus. The  
$s_2$ and $s_{A-2}$ are the spin projections of the nucleon ``2" and ``$A-2$" nucleus respectively.  

In the 2N SRC model we assume that the center of mass momentum of the NN SRC is small which justifies 
the approximation:
\begin{equation}
G(p_{NN}) = \sum\limits_{s_{NN}} \chi_{NN}(p_{NN}, s_{NN})\chi^\dagger_{NN}(p_{NN},s_{NN}),
\label{GNN}
\end{equation}
where $\chi_{NN}$ is the spin wave function and 
$s_{NN}$ is the projection of the total spin of the NN correlation with the three-momentum,  ${\bf p_{NN}} = {\bf p_A} - {\bf p_{A-2}}$.

Inserting  Eqs.(\ref{aops},\ref{vn_poles},\ref{closurule},\ref{GNN})  in Eq.(\ref{sp:2N_cov})  reduces the latter to the 
NN SRC part of the nuclear spectral function, for which one obtains:
\begin{eqnarray}
S_{A,2N}^{N}(p_1,E_m) & = & \sum\limits_{s_2,s_{A-2},s_{NN},s^\prime_{NN}}
\int \chi^\dagger_A \Gamma_{A\to NN,A-2}^\dagger {\chi_{NN}(p_{NN}, s_{NN})
\chi_{NN}^\dagger (p_{NN},s_{NN}) \over p_{NN}^2 - M_{NN}^2 }  \Gamma^\dagger_{NN\rightarrow NN} \nonumber \\
&& {u(p_1,s_1)\over p_1^2 - M_N^2 }   \delta^3(p_1 + p_{2} + p_{A-2}) \delta(E_m - E_m^{2N}){\bar u(p_1,s_1)
\over p_1^2 - M_N^2 }   \nonumber \\
&& {u(p_2,s_2)\bar u(p_2,s_2) \over 2E_2} \Gamma_{NN\rightarrow NN}
 {\chi_{NN}(p_{NN}, s_{NN})\chi^\dagger_{NN}(p_{NN},s_{NN}) \over p_{NN}^2 - M_{NN}^2 }  \nonumber \\
& &  {\chi_{A-2}(p_{A-2},s_{A-2}) \chi^\dagger_{A-2}(p_{A-2},s_{A-2})\over 2E_{A-2}} \Gamma_{A\to NN,A-2} \chi_A 
{d^3p_{2}\over  (2\pi)^3} {d^3p_{A-2}\over  (2\pi)^3}.
 \end{eqnarray}
 Now we introduce  $A\rightarrow  (NN)+(A-2)$ and  $(NN)\rightarrow N+N$  transition wave functions defined 
 in the rest frame of the nucleus $A$ and the 2N SRC respectively (see e.g. Ref.\cite{ms01}):
 \begin{eqnarray}
 \psi_{NN}(p_1,s_1,p_2,s_2) & =  & {\bar u(p_1,s_1)\bar u(p_2,s_2) \Gamma^{NN\rightarrow NN} \chi_{NN}(p_{NN},s_{NN})\over
 (M_N^2 - p_1^2)\sqrt{2E_2(2\pi)^3}} \nonumber \\
  \psi_{CM}(p_{NN},s_{NN},p_{A-2},s_{A-2}) & =  & {\chi^\dagger_{NN}(p_{NN},s_{NN})\chi^{\dagger}_{A-2}(p_{A-2},s_{A-2}) 
  \Gamma^{A\to NN,A-2} \chi_A  
\over (M_{A-2}^2 - p_1^2)\sqrt{2E_{A-2}(2\pi)^3}}, 
\label{wfs}
 \end{eqnarray}
which allows us to present the 2N SRC part of the nuclear spectral function in the following form:
\begin{eqnarray}
 S_{A,2N}^{N}(p_1,E_m)  & = &  \sum\limits_{s_1,s_2,s_{A-2},s_{NN},s^\prime_{NN}}
\int \psi^\dagger_{CM}(p_{NN},s^\prime_{NN},p_{A-2},s_{A-2})\psi^\dagger_{NN}(p_1,s_1,p_2,s_2)\nonumber \\
& & \psi_{CM}(p_{NN},s_{NN},p_{A-2},s_{A-2})\psi_{NN}(p_1,s_1,p_2,s_2) \nonumber \\
& & \delta^3(p_1 + p_2 + p_{A-2})\delta(E_m - E_m^{2N})
d^3p_2  d^3p_{A-2}.
\end{eqnarray}
Using ${\bf p_{NN}} = - {\bf p_{A-2}}$ and integrating by $d^3 p_2$ through $\delta^3(p_1 + p_2 + p_{A-2})$, as well as 
observing that the sum by $s_{A-2}$ results in $\delta_{s_{NN},s^\prime_{NN}}$ one obtains
\begin{eqnarray}
 S^N_{A,2N}(p_1,E_m) =   \sum\limits_{s_1,s_2,s_{A-2},s_{NN}}\int
\mid \psi_{CM}(p_{NN},s_{NN},s_{A-2})\mid^2 \mid \psi_{NN}(p_1,s_1,p_2,s_2)\mid^2 \nonumber \\
\times\delta(E_m - E_m^{2N}) d^3p_{NN}, 
\end{eqnarray}
where $\bf {p_2} = {\bf p_{NN}}  - {\bf p_1}$. 
The above expression is  simplified further by introducing effective momentum distribution 
of the nucleon in the NN SRC, $n_{NN}$, as well as distribution of the center  of mass momentum of the NN correlation, 
$n_{CM}$, which results in,
\begin{equation}
S^N_{A,2N}(p_1,E_m) = \int n_{CM}(p_{NN}) n_{NN}(p_{rel})\delta(E_m - E_m^{2N}) d^3p_{NN},
\label{Snn_vn}
\end{equation}
where ${\bf p_{rel}} ={{\bf p_1} - {\bf p_2}\over 2}$. 

The normalization of spectral function of NN SRC should be related to the total probability of finding nucleon in such a correlation.
This can be defined from the normalization condition  of Eq.(\ref{bnorm})  which yields:
\begin{equation}
\int S^N_{A,2N}(p_1,E_m) \alpha_1 dE_m  d^3p_1 = n^N_{2N},
\end{equation}
where for 2N SRC model 
$\alpha\equiv \alpha_1 = {M_N - E_m - T_{A-1} + p_{1,z}\over  M_A/A}$ and $n^N_{2N}$ is the norm of the SRC contribution of
nucleon $N$ in the total normalization of the nuclear wave function.

As it follows from Eq.(\ref{Snn_vn}) given the relative and center of mass momentum distributions of the NN correlations we 
can numerically calculate the 2N SRC part of the nuclear spectral function.   Since it is assumed that the center of mass momenta 
of the NN SRCs are small,  for $n_{CM}$ we use the distribution obtained in Ref.\cite{CiofiSimula} through the overlap  of two   
Fermi momentum distributions which results in the 
simple gaussian distribution:
\begin{equation}
n_{CM}(p_{NN}) = N_0(A) e^{-\beta(A) p_{NN}^2}
\label{cmdist}
\end{equation}
where $N_{0}(A)$ and  $\beta(A)$ are parameters.
The relative momentum distribution of the NN SRC, $n_{NN}(p_{rel})$,  can be   modeled according to 
Ref.\cite{newprops,proa2}, where the high momentum strength of the nucleon momentum distribution is predicted to be 
inverse proportional to the relative fraction of the nucleon  in the nucleus.  Such a distribution is in agreement with 
the recently observed dominance of $pn$ SRCs\cite{isosrc,Eip3,Eip4} and can be expressed in the form:
\begin{equation}
n^N_{NN}(p_{rel}) = {a_2(A)\over (2x_N)^\gamma} {n_d(p_{rel})\Theta(p_{rel}-k_{src})\over {M_N- E_m - T_{A-1}\over M_A/A} },
\label{2Nmodel}
\end{equation}
where $x_{N} = N/A$ with $N$ being the number of   protons ~(Z) and neutrons (A-Z) in the nucleus A,  
the parameter $a_2(A)$  is related to the probability of finding 2N SRC in the nucleus, $A$,  relative to the 
deuteron and 
$\gamma$ is a free parameter $\gamma \lesssim 1$.  The $n_d(p_{rel})$ is the  high momentum distribution in the deuteron  
and $k_{src}\gtrsim k_{F}$ is the momentum threshold at which 
NN system with such relative momentum can be considered in the short-range correlation.
The factor ${m_N- E_m - T_{A-1}\over M_A/A}$ is the  generalization of the normalization scheme of \cite{FS87} which enforces  the 
normalization condition of Eq.(\ref{bnorm}).

It is worth mentioning that in nonrelativistic limit with assuming an  equal 2N SRC contributions from proton and neutron:
$n^N_{NN}(p_{rel}) = a_2(A) n_d(p_{rel})$ and  the expression in   Eq.(\ref{Snn_vn}) reduces to the "NN SRC-CM motion"  model of 
Ciofi-Simula\cite{CiofiSimula}.

\subsection{Three Nucleon Short-Range Correlations}
\label{sec5c}

For the  3N SRC model in collinear approximation we consider the covariant amplitude of 
Eq.(\ref{sp:3N_cov})  in which the effective vertex $\hat V_{3N}$ is defined as:
\begin{equation}
\hat V_{3N}  = i \bar a (p_1,s_1) \delta^3(p_1 + p_2 + p_{3})\delta(E_m - E_m^{3N}) a (p_1,s_1),
\end{equation}
where  $\bar a (p_1,s_1)$ and $a(p_1,s_1)$ are  creation and annihilation operators defined in Eq.(\ref{aops}).
 
The magnitude of $E_m^{3N}$ is calculated based on the considered  3N SRC model in which recoil nuclear system 
consists of two fast nucleons and a slow $A-3$ nuclear system  whose excitation energy is neglected. In this case
\begin{equation}
E^{3N}_m = E^{(3)}_{thr} + T_{3} + T_{2} - {p_{1}^2\over 2 M_{A-1}},
 \label{e3n}
 \end{equation}
where $E^{(3)}_{thr}$ is the threshold  energy needed to remove three nucleons from the nucleus. 
The kinetic energies of correlated residual nucleons, $T_2$ and $T_3$ are treated relativistically.

According to VN prescription we take integrations by $dp^0_2$ and $dp^0_{3}$  in   Eq.(\ref{sp:3N_cov}), 
at the positive energy poles of propagators of "2" and "3" particles,  i.e.,
\begin{equation}
{ d^0p_{2,3}  \over p_{2,3}^2 - M_N^2+i\varepsilon}  = -{2\pi i  \over 2 E_{2,3}}\left|_{E_{2,3} = \sqrt{M_N^2 + p_{2,3}^2}}. \right.
\label{poles23}
\end{equation}
Using this and the relations of Eq.(\ref{aops}), as well as assuming the sum rule relations (similar to Eq.(\ref{closurule}))
for   the spinors of 2$^\prime$ intermediate state, from  Eq.(\ref{sp:3N_cov})  one obtains for the 3N SRC contribution to 
the nuclear spectral function:
\begin{eqnarray}
& & S^{N}_{A,3N}(p_1,E_m)  =  \sum\limits_{s_{2^\prime},\tilde s_{2^\prime},s_{2},s_{3}} \int \bar u(k_1,\lambda_1) \bar u(k_2,\lambda_2) 
\bar u (k_3,\lambda_3)  \Gamma^\dagger_{NN\rightarrow NN} 
  {u(p_{2^\prime},s_{2^\prime})\bar u(p_{2^\prime},s_{2^\prime})  \over p_{2^\prime}^2 - M_N^2 } \nonumber \\
& & \ \ \ \ \ \ \ \ \times \ \Gamma^\dagger_{NN\rightarrow NN} {u(p_1,s_1)\over p_1^2 - M_N^2} 
\delta^3(p_1 + p_2 + p_3)\delta(E_m - E_m^{3N}) {\bar u(p_1,s_1)\over p_1^2 - M_N^2} 
 {u(p_2,s_2)\bar u(p_2,s_2)\over 2E_2}\nonumber \\ 
&& \ \ \ \ \ \ \ \ \times \ \Gamma_{NN\rightarrow NN} 
 {u(p_{2^\prime},\tilde s_{2^\prime})\bar u(p_{2^\prime},\tilde s_{2^\prime})\over  p_{2^\prime}^2 - M_N^2 } 
 {u(p_3,s_3)\bar u(p_3,s_3)\over 2E_3}  \Gamma_{NN\rightarrow NN} \nonumber \\
& &  \ \ \ \ \ \ \ \ \times \  u(k_1,\lambda_1) u(k_2,\lambda_2) u (k_3,\lambda_3) 
 {d^3p_{3}\over (2\pi)^3} {d^3p_{2}\over (2\pi)^3}.
 \label{sp:3N_cov_b}
 \end{eqnarray}
Introducing $2N$ SRC wave functions in analogy with Eq.(\ref{wfs}):
\begin{equation}
 \psi_{NN}(p_1,s_1,p_2,s_2;p_{1i},s_{1i},p_{2i},s_{2i}) =  
  {\bar u(p_1,s_1)\bar u(p_2,s_2) \Gamma^{NN\rightarrow NN} u(p_{1i},s_{1i})u(p_{2i},s_{2i})\over
 (M_N^2 - p_1^2)\sqrt{2}\sqrt{2E_2(2\pi)^3}},
\end{equation}
where subscript ``i" indicates incoming nucleons with their spin projections,
Eq.(\ref{sp:3N_cov_b})  can be  expressed as follows:
\begin{eqnarray}
S^{N}_{A,3N}(p_1,E_m) & = & \sum\limits_{s_{2^\prime}, \tilde s_{2^\prime},s_2,s_3}
\int \psi^\dagger_{NN}(p_1,s_1,p_2,s_2;k_1,\lambda_1,p_{2^\prime},s_{2^\prime})
\psi^\dagger_{NN}(p_{2^\prime},s_{2^\prime},p_3,s_3;k_2,\lambda_2,k_3,\lambda_3) \nonumber \\
&& \psi_{NN}(p_1,s_1;p_2,s_2;k_1,\lambda_1,p_{2^\prime},s_{2^\prime})
      \psi_{NN}(p_{2^\prime},\tilde s_{2^\prime};p_3,s_3;k_2,\lambda_2,k_3,\lambda_3)\nonumber \\
 && \delta^3(p_1 + p_2 + p_3)\delta(E_m - E_m^{3N}) d^3p_{3} d^3p_{2}.
 \label{sp:3N_cov_c}
 \end{eqnarray}
 Using the fact that the sum over the polarizations of ``2" and ``3" particles 
 results in $s_2= s_{2^\prime} = \tilde s_{2^\prime}$ and taking the $d^3p_2$ integration through the 
$\delta^3(p_1 + p_2 + p_3)$  function, one obtains:
 \begin{eqnarray}
S^{N}_{A,3N}(p_1,E_m)  & =  & \sum\limits_{s_{1},s_2,s_3}
\int \mid \psi_{NN}((p_{2^\prime},s_{2},p_3,s_3;k_2,\lambda_2,k_3,\lambda_3)\mid^2\nonumber \\
& &\mid \psi_{NN}(p_1,s_1;p_2,s_2;k_1,\lambda_1,p_{2^\prime},s_{2})\mid^2 
 \delta(E_m - E_m^{3N}) d^3p_{3}.
\end{eqnarray}
The above expression can be represented in a more simple form by noticing that the NN correlation wave functions 
depend on their relative momenta and we sum over the final  and average by 
all possible initial polarization configurations. This yields:
\begin{equation}
S^{N}_{A,3N}(p_1,E_m)  = \int n_{NN}(p_{2^\prime,3})\cdot  n_{NN}(p_{12})\delta(E_m - E_m^{3N}) 
d^3p_{3},
\label{S3n_vn}
\end{equation}
where  ${\bf p_{12}} = {{\bf p_1}- {\bf p_2}\over 2} = {\bf p_1} + {{\bf p_3}\over 2}$ and 
${\bf p_{2^\prime,3}} = {{\bf p_{2^\prime}} - {\bf p_3} \over 2} \approx - {\bf p_3}$.  The normalization condition for  the 
3N SRC spectral function is  defined as follows:
\begin{equation}
\int S^{N}_{A,3N}(p_1,E_m) \alpha_1 d^3p_1 dE_m = n^N_{3N},
\end{equation}
where  $n^N_{3N}$ is the  norm of the 3N SRC contributing to 
the total normalization of the nuclear wave function for the given nucleon N.
 
Within the model of  $pn$ dominance of two-nucleon SRCs  one predicts that the 
3N SRCs are generated predominantly through the two sequential short range $pn$ interactions. 
As a result our model of 3N SRCs predicts that the overall probability of finding such correlations is proportional to the 
$a_2(A)^2$. Using the relations similar to   Eq.(\ref{2Nmodel}),  one 
approximates: 
\begin{equation}
n_{NN}(p_{2^\prime,3})\cdot  n_{NN}(p_{12}) = a_2(A,Z)^2 C(A,Z) 
{n_d(p_{2^\prime,3}) n_d(p_{12})\over  {M_N - E_m - T_{A-1}\over M_A/A}}\Theta(p_{2^\prime,3}-k_{src})\Theta(p_{12}-k_{src}),
\label{3N_VNanzats}
\end{equation}
where $k_{src}> k_{F}$ is the relative momentum threshold at which the $NN$ system can be considered as a short-range correlation.
Here $C(A,z)$ is function which accounts for the effects associated with the isospin structure of two-nucleon recoil
system. Namely, in the collinear appoximation  two recoil nucleons emerge with small relative momenta (or invariant mass). 
In Ref.\cite{eheppn2} it was demonstrated that the NN system with small  relative momenta is strongly dominated in the 
isosinglet $pn$ channel.  This situation introduces an additional restriction on the isospin composition of the 3N SRCs, in which 
the recoil NN system predominately consists of a $pn$ pair.    For example, one direct consequence of such dynamics is that high 
momentum neutrons can not be generated in 3N SRCs while protons can.

 \section{ Spectral Function in Light-Front Approximation}
\label{Sec.6}

The nuclear spectral function on the light-front  was first formulated in Ref.\cite{FS88} however  its 
calculation from the first principles is impossible 
due to the lack of the  knowledge of nuclear light-front wave functions. 
The current work uses two assumptions which allow us to obtain calculable LF nuclear spectral functions.
The first assumption is that  the nuclear  mean field contribution to  the light-front  spectral function corresponds to the 
non-relativisitc limit of  the  momentum  and  missing energy of a bound nucleon.  As a result the mean-field part of the LF spectral function 
can be related to the mean-field contribution of conventional nuclear spectral function discussed in the previous section.
The second assumption is that the dynamics of the LF spectral function in the high momentum  domain is defined mainly by the $pn$ interaction 
at short distances. Thus to obtain the calculable spectral function in relativistic domain one  will need only a LF model
for the deuteron wave function at short distances.
 
Before to proceed with the above  approach we first define the kinematic parameters that characterize 
the bound nucleon in the light-front as well as the sum rules  that the light-front nuclear spectral function should satisfy.

In defining the  light-front nuclear spectral function the primary requirement is that it is a  Lorentz boost  invariant function  in the 
direction of the large CM momentum of the nucleus $p_A$.  To satisfy this condition we require that the bound nucleon, $N$,
is   described by a light- cone,  "+", momentum fraction $\alpha_N = A{p_{N+}\over p_{A+}}$,  transverse (to ${\bf p_A}$) momentum $\bf p_{N\perp}$ and 
invariant mass of $\tilde M_N^2 = p_{N-}p_{N+} - p_{N,\perp}^2$.  As it will be shown below the $\tilde M_N$ is related to the excitation energy of 
the residual nucleus.   For the future  derivations  it is useful to present the invariant phase space of bound nucleon, $d^4 {p_N}$,  through these light-cone
variables as follows:
\begin{equation}
d^4 p_{N} = {1\over 2} dp_{N-},dp_{N+}d^2 p_{N,\perp} =  {d\alpha_N\over 2 \alpha_N}d^2p_{N,\perp}d \tilde M_N^2.
\label{phspace}
\end{equation} 

Identifying the  kinematic variables  describing  the bound nucleon, one now defines the light-front nuclear spectral function, 
$P_A(\alpha_N,p_{N,\perp},\tilde M_N^2)$,
as a joint probability of finding bound nucleon in the nucleus with light-cone momentum fraction $\alpha_N$, transverse 
momentum $\bf p_{N,\perp}$   and invariant mass $\tilde M_N^2$.  The normalization 
condition for such spectral functions is defined from the requirements of baryonic number  and total light-cone momentum conservations\cite{FS88,FS87} 
\begin{equation}
\int P_A(\alpha_N,p_{N,\perp},\tilde M_N^2)  {d\alpha_N\over 2 \alpha_N}d^2p_{N,\perp}d \tilde M_N^2 = 1 \ \ 
\mbox{and} \ \  \int \alpha_N P_A(\alpha_N,p_{N,\perp},\tilde M_N^2)  {d\alpha_N\over 2 \alpha_N}d^2p_{N,\perp}d \tilde M_N^2 = 1,
\label{sumrules}
\end{equation}
where the second relation is  exact if one assumes that  the  all momentum in the nucleus is carried by the constituent nucleons.
From Eq.(\ref{sumrules}) one  deduces the relation between the LF  spectral function and light-cone density matrix in the form:
\begin{equation}
\rho_{A}(\alpha_N,p_{N,t}) = \int P_A(\alpha_N,p_{N,\perp},\tilde M_N^2) {1\over 2} d \tilde M_N^2. 
\end{equation}

To proceed with the derivations, similar to the VN approximation, 
we follow the decomposition of Fig.\ref{Fig:spectral_diagram} considering  mean-field, 2N and 3N SRC contributions 
separately. 
In LF  approximation the cut diagrams of Fig.\ref{Fig:spectral_diagram} and Fig.\ref{Fig:3nSpectral} will be evaluated at the positive  light-cone ("-" component) energy poles of the spectator residual system. 
For the mean field contribution the spectator residual system consists of  A-1 nucleus, while  for the SRC   case  these  
are  one or  two  on-energy-shell nucleons  correlated with the bound nucleon in  the 2N and 3N  correlations
as well as $A-2$ and $A-3$ uncorrelated  nuclear systems  respectively.

\subsection{Mean Field Approximation}
\label{sec6a}

To calculate the light-front nuclear spectral function in  mean field approximation one needs in principle to start with Eq.(\ref{eq:Smf}) and proceed by evaluating the integral at the pole of the  "minus" 
component of four-momentum of the $A-1$ residual nucleus.  Such an  integration will  express the 
spectral function through the  unknown light-cone mean-field  wave function of the nucleus. 

We adopt a different approach  in which one uses the fact that  the mean field  nuclear spectral function  is  dominated 
at small momenta and  missing energies of bound nucleon
for which the nonrelativistic limit of light-front approximation is well justified.  
Then we need only  to relate the mean field light-cone spectral function  $P^N_{A,MF}(\alpha_1,p_{1,\perp},\tilde M_N^2)$  to the 
above discussed VN mean-field spectral function, $S^N_{A,MF}(E_m,p)$  in nonrelativistic limit\footnote{Note that hereafter we will identify $\alpha_N$ and $p_{N,\perp}$ with $\alpha_1$ and $p_{1,\perp}$ respectively, giving the 
subscript "1" to the bound nucleon.}.
The relation between  $P^N_{A,MF}$ and $S^N_{A,MF}(E_m,p)$ can be found by using
the normalization condition: 
\begin{equation}
\int P^{N}_{A,MF}(\alpha_1,p_{1,\perp},\tilde M_N^2) {d\alpha_1\over 2 \alpha_1}d^2p_{1,\perp}d \tilde M_N^2 = \int S^{N}_{A,MF}(E_m,p)dE_m d^3p_1 = n^{N}_{MF},
\label{mfnorms}
\end{equation}
where we need to relate the LF phase space to $dE_m d^3p$. For this,
we use the relation between the total energy of $A-1$ nucleus and missing energy $E_m$ in mean field approximation:
\begin{equation}
E_{A-1} = \sqrt{M_{A-1}^2 + p_{A-1}^2} = M_A - M_N + E_m + {p_1^2\over 2 M^0_{A-1}},
\label{EA-1}
\end{equation}
where the last part of the equation follows from  the definition of the missing energy $E_m$ (Eq.(\ref{EM})) which is inherently nonrelativistic.
The mass $M^{0}_{A-1}$ represents the ground state mass of the residual $A-1$ system.
With the above equation one obtains for $\alpha_1$:
\begin{equation}
\alpha_1 = A - {E_{A-1} - p_{1,z}\over M_A/A}
\label{alpha_mf}
\end{equation}
and for $\tilde M_N^2$:
\begin{equation}
\tilde M_N^2   =  {\alpha_1\over A}\left (M_A^2 - {M_{A-1}^2 + p_{1,\perp}^2\over (A-\alpha)/A}\right) - p_{1,\perp}^2.
\label{mn_mf}
\end{equation}
These relations allow us to relate:
\begin{equation}
d\tilde M_N^2 d\alpha_1 = 2\alpha d E_m d p_{1,z}.
\end{equation}
This,  together with Eq.(\ref{phspace}) results in  $d^4p_1 = {d\alpha_1\over 2 \alpha_1}d^2p_{1,\perp}d \tilde M_N^2 = d E_m d^3{p_1}$,
which substituting  in  Eq.(\ref{mfnorms}) allows us to obtain for the mean-field approximation
\begin{equation} 
 P^{N}_{A,MF}(\alpha_1,p_{1,\perp},\tilde M_N^2) =  S^{N}_{A,MF}(E_m,p_1),
 \end{equation}
 where $\alpha_1$ and $\tilde M_N^2$ are expressed through $E_m$ and ${\bf p_1}$ according to Eqs.(\ref{EA-1}), (\ref{alpha_mf}) and (\ref{mn_mf}).

\subsection{Two Nucleon Short Range Correlations}
\label{sec6b}

To calculate 2N SRC contribution to the light-front spectral function $P_{A,2N}^{N}(\alpha_1,p_{1,\perp},\tilde M_N^2)$ 
we start with  Eq.(\ref{sp:2N_cov}), with the vertex operator  defined as follows
(see also \cite{FSS15}):
\begin{equation}
\hat V_{2N} = i \bar a(p_1,s_1) 2\alpha_1^2\delta(\alpha_1 + \alpha_2 + \alpha_{A-2}-A)\delta^2(p_{1,\perp}+p_{2,\perp} + p_{A-2,\perp})
\delta(\tilde M_N^2 - \tilde M_{N}^{(2N),2})a(p_1,s_1),
\end{equation}
where $(\alpha_2,p_{2,\perp})$, ($\alpha_{A-2},p_{A-2,\perp}$) are light-cone momentum fractions and transverse momenta 
of correlated second nucleon and residual (A-2) nucleus.
In the considered 2N SRC model: 
\begin{eqnarray}
& & \tilde M_{N}^{(2N),2} = p_{1+}(p_{A-} - p_{2-} - p_{A-2,-}) - p_{1,\perp}^2 = \nonumber \\
& & \ \ \ \ \ \ \ \ \ \ {\alpha_1\over A}\left( M_A^2 - A{M_N^2 + (p_{A-2,\perp} - p_{1\perp})^2\over A-\alpha_1 - \alpha_{A-2}} - 
A{M_{A-2}^2 + p_{A-2,\perp}^2\over \alpha_{A-2}}\right) - p_{1\perp}^2.
\end{eqnarray}
To proceed, in Eq.(\ref{sp:2N_cov}) we treat the nucleon, ``2",  and  residual nucleus, ``A-2", on light-cone energy shells. 
This is achieved by integrating  their respective  $``-"$ components through the pole value of the propagators, provided that their 
$``+"$ components 
are large and positive. For the ``2" nucleon the integration is performed  as follows:
\begin{eqnarray}
{d^4 p_2\over p_2^2 - M_N^2 + i\epsilon} = {{1\over 2}dp_{2-}dp_{2+} d^2 p_{2,\perp}\over p_{2+}p_{2-} - p_{2,\perp}^2 + M_N^2  + i\epsilon} = 
{{1\over 2}dp_{2-}dp_{2+} d^2 p_{2,\perp}\over p_{2+}(p_{2-} - {p_{2,\perp}^2 + M_N^2\over p_{2+}}  + i\epsilon)} \nonumber \\
= - i\pi  {d\alpha_2\over \alpha_2}d^2p_{2,\perp}\left|_{p_{2-} = {p_{2,\perp}^2 + M_N^2\over p_{2+}}}\right. .
\end{eqnarray}
Similarly for the ``A-2" residual nucleus:
\begin{equation}
{d^4 p_{A-2}\over p_{A-2}^2 - M_{A-2}^2 + i\epsilon}  = - i\pi  {d\alpha_{A-2}\over \alpha_{A-2}}d^2p_{A-2,\perp}\left|_{p_{A-2,-} = {p_{A-2,\perp}^2 + M_{A-2}^2\over p_{A-2,+}}}\right. .
\end{equation}
Note that the above integrations project the intermediate state to the positive light-cone energy state thus excluding the contribution from 
the Z - graph of Fig.\ref{Fig:R_diagram}(c).  The $Z$ diagram in this scheme will be suppressed by the inverse power of large $"+"$ component of 
the nucleon's four-momentum (see e.g. \cite{Weinberg}). With the diminished contribution from the $Z$ graph the Eq.(\ref{sp:2N_cov}) will result in 
the  light-front spectral function, $P^N_{A,2N}(\alpha_1,p_{1,\perp},\tilde M_N^2)$  of the 2N SRC.

The on-shell conditions for the nucleon $``2"$ and residual nucleus $``A-2"$ allows to use the sum rule relations:
\begin{equation}
\sh p_2 + M_N = \sum\limits_{s2}u(p_2,s_2)\bar u(p_2,s_2) \ \ \mbox{and} \ \  G(p_{A-2}) = \sum\limits_{s_{A-2}}\chi_{A-2}(p_{A-2},s_{A-2})\chi^\dagger_{A-2}(p_{A-2},s_{A-2}).
\end{equation}
Using also the  non relativistic limit for the center of mass motion of 2N SRC, $k_{cm}\ll M_{NN}$ (for $k_{cm}$ see below),  
we approximate:
\begin{equation}
G(p_{NN},s_{NN}) \approx  \sum\limits_{s_{NN}} \chi_{NN}(p_{NN},s_{NN}) \chi_{NN}^\dagger(p_{NN},s_{NN}),
\end{equation}
where $\chi_{NN}$ is the spin wave function of the C.M. part of the 2N SRC.
Using also the relations $a(p_1,s_1)(\sh p_1 + M_N) = \bar u(p_1,s_1)$ and $(\sh p_1 + M_N)\bar a (p_1,s_1) = u(p_1,s_1)$ for the 
light-front  spectral function,  from Eq.(\ref{sp:2N_cov}) one obtains:
\begin{eqnarray}
& & P_{A,2N}^{N}(\alpha_1,p_{1,\perp},\tilde M_N^2)   =  
 \hskip -12pt  \sum\limits_{s_2,s_{NN},s_{A-2}}\int \chi^\dagger_A  
\Gamma_{A\to NN,A-2}^\dagger \chi_{A-2}(p_{A-2},s_{A-2}) \nonumber \\
& & 
 \ \ \ \ \ \ \ \ \ \  \times \ { \chi_{NN}(p_{NN},s_{NN}) \chi_{NN}^\dagger(p_{NN},s_{NN})  \over p_{NN}^2 - M_{NN}^2}
\Gamma^\dagger_{NN\rightarrow NN} 
{u(p_1,s_1)u(p_2,s_2)\over p_1^2 - M_N^2}  \nonumber \\
& &   \ \ \  \times 
 \left[2\alpha_1^2\delta(\alpha_1+ \alpha_2 + \alpha_{A-2}-A)\delta^2(p_{1,\perp}+p_{2,\perp} + p_{A-2,\perp})
\delta(\tilde M_N^2 - \tilde M_{N}^{(2N),2})\right]
{\bar u(p_1,s_1)\bar u(p_2,s_2)\over p_1^2 - M_N^2 }\nonumber \\ 
& &  \ \ \  \ \ \ \ \ \ \ \times  \ \Gamma_{NN\rightarrow NN}
{ \chi_{NN}(p_{NN},s_{NN}) \chi_{NN}^\dagger(p_{NN},s_{NN})  \over p_{NN}^2 - M_{NN}^2}
\chi^\dagger_{A-2}(p_{A-2},s_{A-2}) \Gamma_{A,NN,A-2}\chi_A \nonumber \\
& &   \ \ \ \ \ \ \ \ \ \ \times \ {d\alpha_2\over \alpha_2}{d^2p_{2,\perp}\over 2(2\pi)^3}
{d\alpha_{A-2}\over \alpha_{A-2}}{d^2p_{A-2,\perp}\over 2(2\pi)^3}.
\label{lc2}
\end{eqnarray} 

\medskip

\noindent{\bf Light-front  wave function of the NN SRC:} 
We now focus on the following combination in Eq.(\ref{lc2}):
\begin{equation}
{\bar u(p_1,s_1)\bar u(p_2,s_2)\over p_1^2 - M_N^2 } \Gamma_{NN\rightarrow NN} \cdot \chi_{NN}(p_{NN},s_{NN}),
\label{2NSRCpart}
\end{equation}
which enters in Eq.(\ref{lc2}) in a direct and complex-conjugated form.  
For the propagator in Eq.(\ref{2NSRCpart}) using light-cone momentum and energy   conservation 
at the $\Gamma_{NN\rightarrow NN}$  vertex  one obtains:
\begin{eqnarray}
p_1^2 - M_N^2 = (p_{NN} - p_2)^2 - M_N^2 = (p_{NN,+}-p_{2,+})\left((p_{NN,-}-p_{2,-}) - {M_N^2 + p_{1\perp}^2\over (p_{NN,+}-p_{2,+})}\right) 
\nonumber \\
= \alpha_1 (p_{A+}/A) \left( {M_{NN}^2  + p_{NN,\perp}^2\over \alpha_{NN}p_{A+}/A} - 
{M_N^2 + p_{2,\perp}^2\over \alpha_{2}p_{A+}/A} - {M_N^2 + p_{1,\perp}^2\over \alpha_{1}p_{A+}/A}\right),
\label{LCprop1}
\end{eqnarray}
where in the last part of the equation we used the on-shell conditions for the nucleon ``2"  ($p_{2-} = {M_N^2 + p_{2,\perp}^2\over p_{2+}}$) and the  condition,  $k_{cm}\ll M_{NN}$, which  justifies 
the approximation, $p_{NN-} \approx {M_{NN}^2 + p_{NN,\perp}^2\over p_{NN+}}$.
Eq.(\ref{LCprop1})  can be further simplified using relations 
$\alpha_1 + \alpha_2 = \alpha_{NN}$ and $p_{1\perp} + p_{2\perp} = p_{NN,\perp}$  yielding:
\begin{equation}
p_1^2 - M_N^2 = {\alpha_1\over \alpha_{NN}}\left(m_{NN}^2 - {\alpha_{NN}^2\over \alpha_1\alpha_2}\left[M_N^2 + (p_{1\perp}^2 - {\alpha_1\over \alpha_{NN}}p_{NN,\perp})^2\right]\right).
\end{equation}
The above propagator can be completely expressed though the internal momenta of the $NN$ system by introducing the  momentum fraction of 
the 2N SRC carried by the nucleon ``1", $\beta_1$, and the relative transverse momentum $k_\perp$ as follows:
\begin{equation}
\beta_1 = 2- \beta_2 = {2\alpha_1\over \alpha_{NN}} \ \ \mbox{and} \ \ k_{1,\perp}  = p_{1,\perp} - {\beta_1\over 2}p_{NN,\perp}.
\end{equation}
With these definitions Eq.(\ref{2NSRCpart}) can be written as:
\begin{equation}
 {\bar u(p_1,s_1)\bar u(p_2,s_2)\over 
 \beta_1 {1\over 2}[ M_{NN}^2 - {4\over \beta_1(2-\beta_1)}(M_N^2 + k_{1,\perp^2})]} 
 \Gamma_{NN\rightarrow NN} \cdot \chi_{NN}(p_{NN},s_{NN}),
\label{2NSRCpart2}
\end{equation}
where one observes that the term in the denominator which is subtracted from $M_{NN}^2$  represents the invariant energy of the $NN$ 
system, $s_{NN}$. This allows us to introduce relative momentum, $k$ in the $NN$ system, which is invariant with respect to the Lorentz  boost in the $\bf p_{NN}$ direction, in the form:
\begin{equation}
s_{NN} = {4\over \beta_1(2-\beta_1)}(M_N^2 + k_{1,\perp}^2) = 4(M_N^2 + k_1^2).
\label{klc}
\end{equation}
The above defined relative momentum, $k_1$, will be used to set a  momentum scale for  the  2N SRCs, requiring 
$k_1\ge k_{src}$ similar to Eq.(\ref{Snn_vn}).

The expression in  Eq.(\ref{2NSRCpart2}) can be presented in a more compact form if one introduces the light-cone wave function of the $NN$ SRC\cite{FS81,FSS15}
in the form:
\begin{equation}
\psi^{s_{NN}}_{2N}(\beta_{1},k_{1,\perp},s_1,s_2) = -{1\over  \sqrt{2(2\pi)^3}} {\bar u(p_1,s_1)\bar u(p_2,s_2)
\Gamma_{NN\rightarrow NN} \cdot \chi_{NN}(p_{NN},s_{NN})\over {1\over 2}[ M_{NN}^2 - 4(M_N^2 + k_1^2)]},
\label{srclcwf}
\end{equation}
where $\chi_{NN}$ represents the  spin wave function of the NN pair emerging from the nuclear vertex $\Gamma_{A,NN,A-2}$.
With this definition for Eq.(\ref{2NSRCpart}) one obtains:
\begin{equation}
{\bar u(p_1,s_1)\bar u(p_2,s_2)\over p_1^2 - M_N^2 } \Gamma_{NN\rightarrow NN} \cdot \chi_{NN}(p_{NN},s_{NN}) = 
-  {\sqrt{2(2\pi)^3}\over \beta_1} \psi^{s_{NN}}_{2N}(\beta_1,k_{1,\perp},s_1,s_2). 
\label{SRCprop}
\end{equation}

\medskip
\medskip

\noindent{\bf Light-front wave function of the NN - (A-2)  System:}
Next we consider the combination that defines the wave function of  ``NN - (A-2)" system that will describe the motion of the 
center of mass of the NN correlation:
\begin{equation}
{ \chi_{NN}^\dagger(p_{NN},s_{NN}) \chi^\dagger_{A-2}(p_{A-2},s_{A-2}) \over p_{NN}^2 - M_{NN}^2}
 \Gamma_{A\rightarrow NN,A-2}\chi_A.
\label{NNA-2}
\end{equation}
We first elaborate the propagator of the $NN$ system using the on-shell conditions for the initial $A$ and residual $A-2$ nucleus:
\begin{equation}
p_{NN}^2 - M_{NN}^2 = p_{NN+}(p_{A-} - p_{A-2,-} - {M_{NN}^2 + p_{NN,\perp}^2\over p_{NN,+}}) = 
 {\alpha_{NN}\over A}(M_A^2 - s_{NN,A-2}),
\label{propNN}
\end{equation}
were in the last part of the equation we used the on-energy shell relations $p_{A-} =  {M_A^2 + p_{A,\perp}^2\over p_{A,+}}$, $p_{A-2,-} = {M_{A-2}^2 + p_{A-2,\perp}^2\over p_{A-2,+}}$,  the definition, $\alpha_{NN} = {A p_{NN,+}\over p_{A+}}$ as well as  introduced
the invariant energy of the $NN$- $A-2$ system as follows:
\begin{equation}
s_{NN,A-2} = A^2{\left[ M_{NN}^2 + {\alpha_{NN}\over A}(M_{A-2}^2 - M_{NN}^2) + (p_{NN,\perp} - {\alpha_{NN}\over A}p_{A,\perp})^2\right]
\over \alpha_{NN}(A-\alpha_{NN})}.
\end{equation}
This invariant energy can be used to estimate the relative three-momentum in the $NN-(A-2)$ system:
\begin{equation}
k_{CM} = {\sqrt{(s_{NN,A-2} - (M_{NN} + M_{A-2})^2)(s_{NN,A-2} - (M_{NN} - M_{A-2})^2)}\over 2\sqrt{s_{NN,A-2}}},
\end{equation}
where $k_{CM,\perp}\equiv k_{NN,\perp} = p_{NN,\perp}$.  Note that $\bf k_{CM}$  can be used to calculate the 
light-cone momentum fraction of the $NN$ pair as follows:
\begin{equation}
\alpha_{NN} = {A(E_{NN} + k_{CM}^z)\over E_{NN} + E_{A-2}},
\end{equation}
where $E_{NN} = \sqrt{M_{NN}^2  + k_{CM}^2}$ and $E_{A-2} = \sqrt{M_{A-2}^2 + k_{CM}^2}$.

With above definitions, similar to Eq.(\ref{srclcwf})  one introduces light-front wave function of $NN-(A-2)$ system:
\begin{equation}
\psi_{CM}(\alpha_{NN},k_{NN,\perp}) = - {1\over \sqrt{{A-2\over 2}}} {1\over \sqrt{2 (2\pi)^3}} 
{  \chi_{NN}^\dagger(p_{NN},s_{NN})  \chi^\dagger_{A-2}(p_{A-2},s_{A-2}) \Gamma_{A\rightarrow NN,A-2}\chi^{s_A}_A
\over {2\over A}\left[ M_{A}^2 - s_{NN,A-2}(k_{CM})\right]},
\label{cmlcwf}
\end{equation}
where for simplicity we suppress the spin notations of $NN$, $A-2$ systems  and the $A$ nucleus.
The coefficients in the above equation are chosen such that in nonrelativistic limit, $k_{CM}\ll M_{NN}$:
\begin{equation}
\psi_{CM}(\alpha_{NN},k_{NN,\perp}) \approx \psi^{NR}_{CM}(k_{CM}) \cdot 2M_N.
\end{equation}
Substituting Eqs.(\ref{cmlcwf}) and (\ref{propNN}) in  Eq.(\ref{NNA-2}) one obtains:
\begin{equation}
{ \chi_{NN}^\dagger(p_{NN},s_{NN}) \chi^\dagger_{A-2}(p_{A-2},s_{A-2}) \over p_{NN}^2 - M_{NN}^2}
 \Gamma_{A,NN,A-2}\chi_A  =  -{\sqrt{{A-2\over 2}}\sqrt{2 (2\pi)^3}\over \alpha_{NN}/2}\psi_{CM}(\alpha_{NN},k_{NN,\perp}).
\label{CMprop}
\end{equation}
\medskip
\medskip
Using Eqs.(\ref{SRCprop}) and (\ref{CMprop}) in Eq.(\ref{lc2}) and restoring  spin indices, for the light-front spectral function 
one obtains:
\begin{eqnarray}
& & P_{A,2N}^{N}(\alpha_1,p_{1,\perp},s_1,\tilde M_N^2,s_A)  =  {A-2\over 2} \sum\limits_{s_2,s_{NN},s_{A-2}}\int 
\psi^{s_A,\dagger}_{CM}(\alpha_{NN},k_{NN,\perp},s_{NN},s_{A-2})\nonumber \\
& &  \times \ \psi^{s_{NN},\dagger}_{NN}(\beta_{1},k_{\perp},s_{1},s_2) 
\left[2\delta(\alpha_1+ \alpha_2 + \alpha_{A-2}-A)\delta^2(p_{1,\perp}+p_{2,\perp} + p_{A-2,\perp})
\delta(\tilde M_N^2 - \tilde M_{N}^{(2N),2})\right]\nonumber \\
& & \ \ \ \times \ \psi^{s_A}_{CM}(\alpha_{NN},k_{NN,\perp},s_{NN},s_{A-2})
\psi^{s_{NN}}_{NN}(\beta_{1},k_{\perp},s_{1},s_2){d\beta_2\over \beta_2}d^2p_{2,\perp}
{d\alpha_{A-2}\over \alpha_{A-2}}d^2p_{A-2,\perp}.
\end{eqnarray}

For spin averaged spectral function the above equation can be  simplified further by 
introducing spin averaged density matrices for the 2N SRC similar to Ref\cite{FS81,FSS15} :
\begin{equation}
\rho^{N}_{NN}(\beta_1,k_{1,\perp}) = {1\over 2}{1\over 2 s_{NN}+1}\sum\limits_{s_{NN},s_1,s_2}
{\mid \psi^{s_{NN}}_{NN}(\beta_1,k_{1,\perp},s_1,s_2)\mid^2\over 2-\beta_1},
\label{denmatrix}
\end{equation}
and the  new density matrix for the center of mass motion of the 2N SRC:
\begin{equation}
\rho_{CM}(\alpha_{NN},k_{NN,\perp}) = {1\over 2}{A-2\over 2s_{A} + 1}\sum\limits_{s_{NN},s_{A-2}} 
{\mid\psi^{s_A}_{CM}(\alpha_{NN},k_{NN,\perp},s_{NN},s_{A-2})\mid^2\over A - \alpha_{NN}}.
\end{equation}
With the above definitions, for the unpolarized light-front spectral function one obtains:
\begin{eqnarray}
P_{A,2N}^{N}(\alpha_1,p_{1,\perp},\tilde M_N^2) = {1\over 2}\int \rho^{N}_{NN}(\beta_1, k_{1,\perp})\rho_{CM}(\alpha_{NN},k_{NN,\perp}) 
2\delta(\alpha_1+ \alpha_2 - \alpha_{NN}) \nonumber \\ 
\delta^2(p_{1,\perp}+p_{2,\perp} - p_{NN,\perp}) 
\delta(\tilde M_N^2 - \tilde M_{N}^{(2N),2}) d\beta_2 d^2p_{2,\perp}
d\alpha_{NN}d^2p_{NN,\perp}.
\label{P2Nsrc}
\end{eqnarray} 
The normalization conditions for the above introduced density matrices are 
defined from the sum-rule conditions of Eq.(\ref{sumrules}).
For the density matrices of  NN SRC, $\rho^{N}_{NN}$,  the  normalization conditions to satisfy the baryonic  and momentum sum rules\cite{FS81} 
yield:
\begin{equation}
\int \rho^{N}_{NN}(\beta,k_\perp){d\beta\over \beta}d^2k_\perp = \int \rho^N_{NN}(\beta,k_\perp)\beta {d\beta\over \beta}d^2k_\perp = n^N_{2N},
\end{equation}
where $n^N_{2N}$ is the contribution of the 2N SRCs   to the total norm of the spectral function.
Similar to Eq.(\ref{2Nmodel}) we can model the light-front density matrix of the 2N SRC  through the high momentum 
part of the  light-front density matrix of the deuteron $\rho_d(\beta_1,k_{1,\perp})$ in the form:
\begin{equation}
\rho^N_{NN}(\beta_1,k_{1,\perp}) = {a_2(A)\over (2x_N)^\gamma} \rho_d(\beta_1,k_{1,\perp})\Theta(k_1-k_{src}),
\end{equation}
where $k_1$ is defined in Eq.(\ref{klc}). In the second part of the current work\cite{multisrcII} we will discuss the specific models for 
$\rho_d(\beta_1,k_{1,\perp})$ which will allow 
us to perform numerical estimates.

For the light-front density function of the center of mass motion the conditions of Eq.(\ref{sumrules}) require the following 
normalization relations:
\begin{equation}
\int \rho_{CM}(\alpha_{NN},k_{NN,\perp}){d\alpha_{NN}\over \alpha_{NN}} d^2k_{NN,\perp} = 1 \ \ \mbox{and} \ \ 
 \int \rho_{CM}(\alpha_{NN},k_{NN,\perp})\alpha_{NN}{d\alpha_{NN}\over \alpha_{NN}} d^2k_{NN,\perp} = 2.
\label{CMnorms}
\end{equation}
Since in the considered 2N SRC model the CM motion is nonrelativistc ($k_{NN}\ll 2M_{N}$),  we can use the momentum 
distribution used in VN approximation (Eq.(\ref{cmdist})) which can be related to $\rho_{CM}$ as follows:
\begin{equation}
\rho_{CM}(\alpha_{NN},k_{NN,\perp})  = {E_{NN} E_{A-2}\over (E_{NN} + E_{A-2})/A} {n_{CM}(k_{CM})\over A-\alpha_{NN}}\approx 2M_{N}n_{CM}(k_{CM}),
\end{equation}
where $n_{CM}$ is defined in Eq.(\ref{cmdist}).
Note that for the "middle" form of the $\rho_{CM}$ the first normalization condition of Eq.(\ref{CMnorms}) is exact while the 
second one is approximate satisfying it only in the nonrelativistic limit (last part of the equation).   

Finally, it is worth mentioning that in the nonrelativistic limit of the density matrix of 2N SRC, $\rho(\beta_1,k_{1,\perp}) \approx n_{NN}(p_1)\cdot M_N$
and  Eq.(\ref{P2Nsrc}), similar to VN approximation,  reduces to the SRC  model of spectral function of Ref.\cite{CiofiSimula}.

\subsection{Three-Nucleon Short-Range Correlation Model}
\label{sec6C}
To calculate the contribution of the 3N SRCs to the light-front spectral function we adopt the collinear approach discussed in 
Sec.\ref{sec4c}. In this approach the assumption of the total momentum of 3N SRCs  being  negligible imposes 
several kinematic restrictions on the light-cone momenta of nucleons in the correlation.
The vertex operator, $\hat V_{3N}$ entering in Eq.(\ref{sp:3N_cov}) takes into account these kinematic restrictions in the 
following form:
\begin{equation}
\hat V_{3N} = i \bar a(p_1,s_1) 2\alpha_1^2\delta(\alpha_1 + \alpha_2 + \alpha_{3})\delta^2(p_{1,\perp}+p_{2,\perp} + p_{3,\perp}) \delta(\tilde M_N^2 - \tilde M_{N}^{(3N),2})a(p_1,s_1),
\end{equation}
where in the considered 3N SRC model:
\begin{equation}
\tilde M_{N}^{(3N),2} = p_{1+}(p_{A-} - p_{2-} - p_{3-} -
p_{A-3,-}) - p_{1,\perp}^2 = {\alpha_1\over 3}\left( M_{3N}^2 - {M_N^2 + p_{2,\perp}^2\over \alpha_{2}/3} - 
{M_N^2 + p_{3,\perp}^2\over \alpha_{3}/3}\right) - p_{1,\perp}^2,
\end{equation}
with the mass of the 3N SRC  defined as:
\begin{equation}
M_{3N}^2 = {3\over A} M_A^2 - 3 {M_{A-3}^2\over \alpha_{A-3}}.
\end{equation}
The following derivation in many ways  similar to the one in the previous section.  We first substitute the vertex function $V_{3N}$ into 
Eq.(\ref{sp:3N_cov}), expressing four-dimensional differentials through the light-cone momenta. Then we estimate  the $dp_{2,-}$ and $dp_{3,-}$ integrals at the pole values of the propagators of ``2" and  ``3" nucleons  corresponding to the 
positive values of their ``+" components, as follows:
\begin{equation}
{d^4p_{2/3}\over p_{2/3}^2 - M_N^2 + i\epsilon} = -i(2\pi) {dp_{2/3,+} d^2 p_{2/3,\perp}\over 2p_{2/3,+}}  = -i\pi {d\alpha_{2/3}\over \alpha_{2/3}} d^2 p_{2/3,\perp}\left |_{p_{2/3-}} = {M_N^2 + p_{2/3,\perp}^2\over p_{2/3+}}.\right.
\end{equation}
The above integrations allow us to use the 
``one-mass-shell" sum rule relations for the numerators of the propagators of "2" and "3" nucleons $\sh p + M_N = \sum\limits_{s} u(p,s)\bar u(p,s)$. 
Using the similar approximate relation for the "2$^\prime$"  propagator which represents the nucleon between consecutive $NN$ interaction vertices 
as well as the properties of creation, $a(p_1,s_1)^\dagger$  and annihilation, $a(p_1,s_1)$ operators, for the 3N SRC light-front spectral function  one obtains from Eq.(\ref{sp:3N_cov}):
\begin{eqnarray}
&&P^N_{A,3N}(\alpha_1,p_{1,\perp},s_1,\tilde M_N) = \sum\limits_{s_2,s_3,s_{2^\prime},\tilde s_{2^\prime}} \int \bar u(k_1) \bar u(k_2)\bar u(k_3) \Gamma^\dagger_{NN\rightarrow NN} 
 {u(p_{2^\prime},\tilde s_{2^\prime}) \bar u(p_{2^\prime},\tilde s_{2^\prime})  \over p_{2^\prime}^2 - M_N^2} 
 \nonumber \\
  &&\times \ 
 \Gamma^\dagger_{NN\rightarrow NN}
 { u(p_1,s_1) \over p_1^2 - M_N^2}  
  u(p_2,s_2)  \left[2\alpha_1^2 \delta(\alpha_1 + \alpha_2  + \alpha_3 - 3) \delta^2(p_{1\perp} + p_{2\perp} + p_{3\perp}) \delta(\tilde M_N^2 - M_N^{3N,2})\right] \nonumber \\
&& \times \  \bar u(p_2,s_2) {\bar u(p_1,s_1) \over p_1^2 - M_N^2} \Gamma_{NN\rightarrow NN} {u(p_{2^\prime}, s_{2^\prime}) \bar u(p_{2^\prime},s_{2^\prime})  \over p_{2^\prime}^2 - M_N^2} 
 u(p_3,s_3) \bar u(p_3,s_3) \Gamma^\dagger_{NN\rightarrow NN}u(k_1) u(k_2) u(k_3)\nonumber \\
& &  \times  \ {d\alpha_2\over \alpha_2} {d^2 p_{2\perp}\over 2(2\pi)^3}{d\alpha_3\over \alpha_3}{d^2 p_{3\perp}\over 2(2\pi)^3},
 \label{3NSpectralLCa}
\end{eqnarray}
where we suppress the spin notations of the initial and final collinear nucleons for simplicity of expressions.
Next we consider the  following combination from the above expression:
\begin{equation}
 {\bar u(p_1,s_1) \bar u(p_2,s_2) \Gamma_{NN\rightarrow NN} u(p_{2^\prime}, s_{2^\prime}) u(k_1)  \over p_{1}^2 - M_N^2}. 
\label{termI}
\end{equation}
Here the denominator, similar to the previous section can be expressed through the relative light-cone momentum variables:
\begin{equation}
p_1^2 - M_N^2 = {\beta_1\over 2}\left[M_{12}^2 - {  4[M_N^2 + (p_{1\perp} - {\beta_1\over 2}p_{12,\perp})^2]\over (2-\beta_1)\beta_1}\right],
\label{denom1}
\end{equation}
where we also  applied the kinematic conditions following from the  collinear approximation:
\begin{eqnarray}
M_{12}^2 & = &  (k_1 + k_2 + k_3 - p_3)^2  \approx M_{3N}^2(1 -  {\alpha_3\over 3}) - 3{M_N^2 + p_{3\perp}^2\over \alpha_3} + M_N^2, \\ \nonumber 
p_{12,\perp} &  \approx  & -p_{3\perp} \ \ \ \ \mbox{and} \ \ \ \beta_1 = {2\alpha_1\over \alpha_{12}} = {2\alpha_1\over 3-\alpha_3},
\end{eqnarray}
with $\alpha_{12} = 3{p_{1+} + p_{2+}\over k_{1+} + k_{2+} +k_{3+} }$.

Using  Eq.(\ref{denom1}) one introduces the light-cone wave function of $NN$ SRC simliar  to Eq.(\ref{srclcwf}):
\begin{equation}
\psi_{2N}(\beta_1,k_{1\perp},s_1,s_2) = -{1\over  \sqrt{2(2\pi)^3}} { \bar u(p_1,s_1)\bar u(p_2,s_2)
\Gamma_{NN\rightarrow NN}  u(p_{2^\prime},s_{2^\prime})u(k_1)\over {1\over 2}[ M_{12}^2 - 4{M_N^2 + k_{1\perp}^2\over \beta_1(2-\beta_1)}]},
\label{NNlcwfa}
\end{equation}
where $\tilde k_{1\perp} = p_{1\perp} - {\beta_1\over 2}p_{12,\perp}$ and we also define  the relative momentum in 
the NN center of mass frame as:
\begin{equation}
\tilde k_1^2 = {M_N^2 + \tilde k_{1\perp}^2\over \beta_1(2-\beta_1)} - M_N^2.
\label{k1}
\end{equation}

With the above  definitions  for Eq.(\ref{termI}) one obtains:
\begin{equation}
{\bar u(p_1,s_1) \bar u(p_2,s_2) \Gamma_{NN\rightarrow NN} u(p_{2^\prime}, s_{2^\prime}) u(k_1)  \over p_{1}^2 - M_N^2} = 
 \sqrt{2(2\pi)^3} {\psi_{2N}(\beta_1,\tilde k_{1\perp},s_1,s_2) \over \beta_1}.
\label{termIf}
\end{equation}

For the second NN SRC contribution in Eq.(\ref{3NSpectralLCa}) we consider the term:
\begin{equation}
 {\bar u(p_{2^\prime},s_{2^\prime}) \bar u(p_3,s_3) \Gamma_{NN\rightarrow NN} u(k_2) u(k_3)  \over p_{2^\prime}^2 - M_N^2}, 
\label{termII}
\end{equation}
for which the denominator similar to  Eq.(\ref{denom1}) can be represented in the form:
\begin{equation}
p_{2^\prime}^2 - M_N^2  = {2-\beta_3\over 2}\left[M_{23}^2 - 4{M_N^2 + p_{3\perp}^2\over \beta_3(2-\beta_3)}\right],
\label{denom23}
\end{equation}
where the  several relations below  follow from the collinear approximation:
\begin{eqnarray}
& & M_{23}^2  = (k_2 +k_3)^2 \approx 4M_N^2; \ \ \  p_{2^\prime\perp}\approx - p_{3\perp};   \nonumber \\  
& & \alpha^\prime_{23} = {p_{2^\prime,+} + p_{3+}\over k_{1+}+k_{2+}+k_{3+}} \approx 2; \ \ \ \beta_{2^\prime}  = \alpha_{23} - \beta_3 \approx 2-\beta_3,
\label{def23}
\end{eqnarray}
and the relative momentum in the NN center of mass frame can be defined as:
\begin{equation}
\tilde k_3^2 = {M_N^2 + p_{3\perp}^2\over \beta_3(2-\beta_3)} - M_N^2.
\label{k3}
\end{equation}

Eqs.(\ref{denom23}) and (\ref{def23}) allow us to use the definition of the $NN$ SRC wave function of Eq.(\ref{NNlcwfa})  with the replacements of $M_{12}\rightarrow M_{23}$,
$\beta_{1} \rightarrow 2-\beta_{3}$, $k_{1\perp} \rightarrow -p_{3\perp}$ to describe the wave function of the second NN correlation.
This results in the following expression for Eq.(\ref{termII}) 
\begin{equation}
 {\bar u(p_{2^\prime},s_{2^\prime}) \bar u(p_3,s_3) \Gamma_{NN\rightarrow NN} u(k_2) u(k_3)  \over p_{2^\prime}^2 - M_N^2} = 
 \sqrt{2(2\pi)^3} {\psi_{2N}(\beta_3,p_{3\perp},s_{2^\perp},s_3) \over 2-\beta_3}.
\label{termIIf}
\end{equation}

Using Eqs.(\ref{termIf}) and (\ref{termIIf}) in  Eq.(\ref{3NSpectralLCa}) for the light-front spectral function one arrives at:
\begin{eqnarray}
&&P^N_{A,3N}(\alpha_1,p_{1,\perp},s_1,\tilde M_N) =\sum\limits_{s_2,s_3,s_{2^\prime},\tilde s_{2^\prime}} \int 
{\psi^\dagger_{NN}(\beta_3,p_{3\perp},\tilde s_{2^\prime},s_3)\over 2-\beta_3}{\psi^\dagger_{NN}(\beta_1,\tilde k_{1\perp}, s_{1},s_2) 
\over \beta_1} \nonumber \\
& &   \ \ \ \ \ \times \left[2\alpha_1^2 \delta(\alpha_1 + \alpha_2  + \alpha_3 - 3) \delta^2(p_{1\perp} + p_{2\perp} + p_{3\perp}) 
\delta(\tilde M_N^2 - M_N^{3N,2})\right] 
\nonumber \\
& & \ \ \ \ \ \times {\psi_{NN}(\beta_1,\tilde k_{1\perp}, s_{1},s_2) \over \beta_1} {\psi_{NN}(\beta_3,p_{3\perp},\tilde s_{2},s_3)\over 2-\beta_3} 
{d\alpha_2\over \alpha_2} d^2 p_{2\perp}{d\alpha_3\over \alpha_3}d^2 p_{3\perp}.
 \label{3NSpectralLCb}
\end{eqnarray}

In the above expression,  the  unpolarized spectral function can be expressed through the light-front density matrices of
Eq.(\ref{denmatrix}) corresponding to the first and second NN correlations, in the form:
\begin{eqnarray}
P^N_{A,3N}(\alpha_1,p_{1,\perp},\tilde M_N)  &  =  & \int {3-\alpha_3\over 2(2-\alpha_3)^2} \rho_{NN}(\beta_3,p_{3\perp}) 
\rho_{NN}(\beta_1,\tilde k_{1\perp}) 
2 \delta(\alpha_1 + \alpha_2  + \alpha_3 - 3)\nonumber \\
& &  \delta^2(p_{1\perp} + p_{2\perp} + p_{3\perp}) 
 \delta(\tilde M_N^2 - M_N^{3N,2})  d\alpha_2 d^2 p_{2\perp}d\alpha_3d^2 p_{3\perp},
 \label{P3Nsrc}
\end{eqnarray}
 where within colinear approximation $\beta_3 = \alpha_3$, $\beta_1 = {2\alpha_2\over 3-\alpha_3}$ as well as $\tilde k_{1\perp} = p_{1\perp} + {\beta_1\over 2} p_{3\perp}$. 
 In the above expression similar to Eq.(\ref{3N_VNanzats}) 
 the product of two density matrices is expressed through the product of high momentum parts of the 
 deuteron density matrices in the form:
\begin{equation}
\rho_{NN}(\beta_3,p_{3\perp}) \rho_{NN}(\beta_1,k_{1\perp})  = a_2(A,Z)^2 C(A,Z) \rho_{d}(\beta_3,p_{3\perp})\Theta(\tilde k_1-k_{src}) \rho_{d}(\beta_1,\tilde k_{1\perp}) \Theta(\tilde k_3-k_{src}),
\label{3N_LFanzats}
\end{equation}
where $\tilde k_1$ and $\tilde k_3$ are defined in Eqs.(\ref{k1}) and (\ref{k3}) respectively.  The factor $C(A,Z)$ is the same as 
in the case of 3N SRCs within VA approximation.

 \section{Summary}
 \label{Sec.7}
Based on the NN short-range correlation picture of high-momentum component of nuclear wave function we developed a model 
for the nuclear spectral functions in  the domain of large momentum and removal energy of bound nucleon in the nucleus.
Our main focus is in treating the relativistic effects which are important for the bound nucleon momenta   exceeding characteristic 
Fermi momentum in the nucleus $k_{F}$.  The relativistic effects in this work are treated based on the effective Feynman 
diagrammatic approach, in which one starts with Lorentz covariant amplitudes reducing them to the nuclear spectral functions that allows to 
trace the relativistic effects entering in these functions.
One of the main ambiguities related to the treatment of the relativistic effects is the account for the vacuum fluctuations  (Z-graphs) which 
ultimately alter the  definition of the spectral function as a probability of finding a bound nucleon in the nucleus with the given 
momentum and removal energy.
We employed two:  virtual nucleon and light-front approaches in treating the relativistic effects.  

The results for the 2N SRC model within VN~(Eq.(\ref{Snn_vn})) and LF~(Eq.(\ref{P2Nsrc})) approximations agree with the 2N SRC (with center of mass motion) model of 
Ref.\cite{CiofiSimula} in the nonrelativistic limit. Our results represent the first attempt to account for the relativistic effects in the domain of 2N SRCs with center of mass motion 
of the NN pair.

We extended both approaches to calculate also the contributions from 
three-nucleon short-range correlations. Derivations in 
this case are based on 
the collinear approach in which one assumes  negligible center of mass momentum for the residual/uncorrelated $(A-3)$ nuclear system. 
The derived spectral functions within  VN~(Eq.(\ref{S3n_vn})) and LF~(Eq.(\ref{P3Nsrc})  approximations represent the first results for 3N SRC 
contribution to the nuclear spectral functions.

The main property of the obtained spectral functions is that to describe them quantitatively in high momentum domain
one needs  only the knowledge  of  the high momentum deuteron 
wave function either in the Lab frame (for VN approximation) or on the light-front (for LF approximation).  
In the follow-up work\cite{multisrcII} we will present the quantitative studies of the properties of nuclear spectral functions based on 
specific models of  the  deuteron wave functions.

\subsection*{Acknowledgments}
We are thankful to J.~Arrington, W.~Boeglin, W.~Cosyn, D.~Day, L.~ Frankfurt, A.~Freese,  O.~Hen, E.~Piasetzky, J.~Ryckebusch, M~Strikman  and L.~Weinstein for numerous discussions and valuable comments during our research on the project.
This work is supported by U.S. Department of Energy grant under contract DE-FG02-01ER41172.



\begin{thebibliography}{16}
\bibitem{goodcomp}This statement is exact for specific component of 
the interacting nucleon spinor, which in electromagnetic interaction is constitutes to the "good" component of 
the nuclear electromagnetic current.
\end{thebibliography}
\end{document}